\documentclass[prd,a4paper,aps,twocolumn,superscriptaddress,longbibliography]{revtex4-2}
\usepackage{amssymb,amsmath,amsfonts,bm}
\usepackage[normalem]{ulem}

\usepackage{graphicx}
\usepackage{epstopdf}
\usepackage{times}
\usepackage{float}
\usepackage{lipsum}
\usepackage{color}
\usepackage{dcolumn}
\usepackage{bm}
\usepackage{graphicx}
\usepackage{subfigure}
\usepackage[colorlinks,linkcolor=blue,anchorcolor=blue,urlcolor=blue,citecolor=blue]{hyperref}
\usepackage{verbatim} 
\usepackage{amsmath} 
\usepackage{tikz}
\usepackage{braket}
\usetikzlibrary{quantikz}

\providecommand{\theoremname}{Theorem}
\newcommand*{\myproofname}{Proof}

\newcommand{\sgn}{ \mathrm{sgn}}

\usepackage{hyperref}
\hypersetup{colorlinks=true, linkcolor=blue, citecolor=red, urlcolor=blue  }

\usepackage{physics}

\begin{document}

\title{Conformally Invariant Brans-Dicke Loop Quantum Cosmology: A Quantum Geometric Model of Linking Theory }

\author{Chun-Yen Lin }
\email{cynlin@ucdavis.edu}

\affiliation{Department of Physics, Soochow University, Taipei 111002, R.O.C.}
\affiliation{Department of Physics, Beijing Normal University, Beijing 100875, China}

\author{ Xiangjing Liu}
\email{liuxj@mail.bnu.edu.cn}

\affiliation{Department of Physics, Beijing Normal University, Beijing 100875, China}

\affiliation{Department of Physics, Southern University of Science and Technology, Shenzhen 518055, China}

\author{ Yongge Ma}
\email{mayg@bnu.edu.cn}

\affiliation{Department of Physics, Beijing Normal University, Beijing 100875, China}

\author{Cong Zhang}
\email{zhang.cong@mail.bnu.edu.cn}

\affiliation{Institut f\"ur Quantengravitation, Friedrich-Alexander Universit\"at Erlangen-N\"urnberg, Staudtstr. 7/B2, 91058 Erlangen, Germany}


\begin{abstract}  

The loop quantization of the conformal Brans-Dicke cosmology is explored in the spatially flat and Bianchi-I setting. The scalar and conformal constraints governing the canonical model are quantized using the loop techniques. The physical Hilbert space of quantum spacetimes satisfying both quantum constraints is then obtained by incorporating the quantum geometric features. The Schr\"odinger cosmic evolutions are derived with the relational Heisenberg observables describing the dynamical degrees of freedom with respect to the chosen reference degrees of freedom, with the latter providing the physical coordinates for the spatial hypersurfaces and the conformal scales. We show that the emerging Schr\"odinger theories contain not only the loop quantum cosmology of general relativity, but also that of the so-called shape dynamics. The exact dictionary between the two theories is achieved via the underlying physical Hilbert space possessing the additional (loop-corrected) conformal symmetry. 
 \end{abstract}

\maketitle

\section{Introduction}

Loop quantum gravity (LQG) is a non-perturbative, background-independent approach to the fundamental theory of quantum gravity \cite{ashtekar2004background,han2007fundamental}; it strives to provide a concrete prescription of fully dynamical spacetime in terms of the Planck-scale quantum geometry. While the full theory is still developing with many challenges, its cosmological models--  collectively called loop quantum cosmology (LQC)  \cite{ashtekar2003mathematical,ashtekar2011loop}-- have consistently demonstrated prospects of resolving crucial problems in cosmology, by the effects of the discrete spatial geometry which is the foundation of LQG. As an important example, one universal feature of cosmic evolutions in LQC is the replacement of the big bang singularity with a non-singular quantum bounce\cite{ashtekar2006quantum1,ashtekar2006quantum} due to the dominant quantum geometry effects near the Planck regime. In reverse, by testing the effects of the key ingredients of LQG in the cosmological setting, the LQC models also provide useful guidance for the next steps of the full theory’s development.

Recently, the arena of this interplay between LQG and LQC has been extended to the scalar-tensor (ST) theories of gravity, which describe the vacuum universe with an additional scalar field non-minimally coupled to the spacetime geometry. The Brans-Dicke (BD) theory is known as the simplest version of ST theories by using the scalar field as the dynamical gravitational `constant' \cite{brans1961mach} to incorporate Mach's principle. Under the choice of the BD coupling parameter $\omega=0$ or $\omega=-3/2$, the BD theory has been shown to be equivalent to respectively the metric or Palatini formulation of $f(R)$ gravity \cite{sotiriou2010f}, which is making progress in the problems of dark energy and dark matter \cite{bohmer2008dark,artymowski2014inflation, yadav2019dark}. 

The loop quantization procedure has been successfully applied to 
the scalar-tensor theories \cite{zhang2011nonperturbative}, the BD theory with the above coupling-parameter values \cite{zhang2012loop}, and also the metric $f(\mathcal{R})$ theories\cite{zhang2011extension,zhang2011loop}. These works explore the interesting new physics of LQG in the context of Mach’s principle and scalar-tensor cosmology. In the cosmological models of BD LQC with $\omega\neq-3/2$, the dynamics has been obtained in detail with interesting implications of the emergent time variable from the gravitational scalar field~\cite{zhang2013loop,Song:2020pqm}.

However, there remained a singular case of the BD theory that has not been studied with the loop quantization; it is the case of the coupling parameter $\omega=-3/2$ and a trivial scalar potential. Under this setting the vacuum BD theory becomes conformally invariant. Following the two motivations below, we study the loop quantum cosmology of the conformal BD theory in this work.

First, the conformal BD theory has been understood to be a minimal augmentation of general relativity (GR)— endowed with an additional scalar field and the associated conformal gauge symmetry for the correct gravitational degrees of freedom \cite{sotiriou2010f,mercati2014shape}.
 Remarkably, under the various symplectic reductions of the conformal and re-foliation symmetries, the conformal BD theory yields the Hamiltonian dynamics of GR in constant-mean-curvature foliations \cite{mercati2014shape}, unimodular gravity and shape dynamics \cite{ mercati2014shape,gielen2018gravity}. These existing theories of gravity are thus the various representations for the conformal BD theory as the ``linking theory”, obtained with the various choices in an augmented notion of reference frames. In particular, the GR is recovered in the frames with a constant scalar-field gauge-fixing condition, while the shape dynamics can be obtained in the frames with instead a constant conjugate-momentum condition on the scalar field. The study of LQC of the conformal BD theory is thus a meaningful first step, toward a quantum linking theory for the seemingly distinct canonical theories of LQG, such as the GR LQG \cite{ashtekar2004background,han2007fundamental} and the shape dynamics LQG \cite{smolin2014linking}.

Our second motivation lies in the problem of renormalization in quantum gravity. In contrast to the case of the standard model of matter, it has been widely demonstrated that generic effective theories of quantum gravity change drastically with the energy scales. These highly non-trivial renormalization flows in the theory-parameter space pose a serious challenge for the goal of finding a fundamental Planck-scale theory of gravity, that can be shown to give the correct low-energy effective theory of GR (with proper corrections). Nevertheless, a certain class of $f(R)$ theories, treated as the low-energy effective theories, has been observed to have significantly better-behaved renormalization flows, which may lead to a well-defined high-energy limit \cite{smolin2014linking,ohta2016renormalization,eichhorn2018asymptotically}. Especially, in what is called the asymptotic safety scenario \cite{eichhorn2018asymptotically}, there exists a non-trivial fixed point for the high-energy end of the flow, representing the absolute microscopic theory of quantum gravity. Plausibly, such a scale-invariant fixed point may be given by a conformal scalar-tensor theory as the high-energy limit of the effective theory.

For LQG as a fundamental Planck-scale theory, the renormalization problem lies in the downstream direction: what precise form should LQG adopt to ensure that it gives the observed low-energy effective theory, across such a large gap of energy scales? The above asymptotic safety scenario has motivated us to study the LQG quantization of the conformal BD theory as the simplest conformal scalar-tensor theory. This way, we may explore the possibility of the running effective theory first being trapped in the fixed-point neighborhood of the conformal BD theory, before finally being driven away by the accumulated microscopic quantum fluctuations encoding the invariance-breaking LQG quantum geometry. In this controlled manner, we may hope for arriving at the low energy effective theory in the form of (corrected) GR, as the perturbed conformal BD by the LQG quantum fluctuations.

Here, we construct a conformal-BD LQC model under the vacuum and spatially flat Bianchi I setting. In order to capture the essential physics concerning the above topics in this model, we will proceed in the following manner. The fully quantum and interacting nature of the metric-scalar system is essential for both the linking theory and the idea of asymptotic safety; Therefore we will maximally maintain the background independence of LQG, constructing the physical Hilbert space by imposing the loop quantized scalar and conformal constraints, upon the kinematic Hilbert space with all the fields quantized in equal footing. The two quantum constraints will be constructed using standard loop quantization techniques, such that each solution state in the physical Hilbert space is endowed with the characteristic quantum geometry as a conformal-BD quantum spacetime. Further, we introduce the mentioned augmented reference frames at the quantum level by applying the quantum reference frames \cite{lin2016quantum,lewandowski2018quantum} approach to our model. Each chosen frame will be specified with the ``reference” quantum degrees of freedom providing the coordinates for the spatial hypersurfaces and the conformal scales, with respect to which the remaining `` dynamical” quantum degrees of freedom may be described as evolving in the time coordinate of the frame. The instantaneous observables of different moments constructed this way would be acting in the timeless physical Hilbert space and related by the Heisenberg evolution, under the given frame.       

In this work, we will construct a conformal BD LQC model with a given physical Hilbert space, in which the instantaneous quantum relational observables are explicitly defined via the various augmented quantum reference frames. Moreover, we will show that the Schr\"odinger theories of both GR LQC and shape-dynamics LQC emerge through the relational observables associated with the respective frames. Further, since the observables for both frames are defined in the same physical Hilbert space of the BD LQC acting as the linking theory, we will be able to compute the correspondence transformation of the wave functions between the two theories. Lastly, we will study the explicit forms of our relational observables, and their Schr\"odinger propagators will be shown to take the (loop-corrected) Fadeev-Popov path integrals of the model. This serves as the foundation for future study on the renormalization of the model, either through the canonical or path integral approach.

\section{  Conformal Brans-Dicke cosmology in connection variables: Bianchi-I setting }

 The original action of four-dimensional vacuum BD theory reads
  \begin{equation}\label{Brans-Dicke Action}
S(g,\phi)=\frac{1}{2}\int _\Sigma\sqrt{-g} d^4x[\phi \mathcal{R}-\frac{\omega}{\phi}(\partial^\mu\phi)(\partial _\mu\phi))],
 \end{equation}
 where we {choose the geometric unit to set $c=1$,  $8\pi G=1$ such that the Planck length reads $\ell_{Pl}=\sqrt{\hbar}$},  $\phi$ is a scalar field, $\mathcal{R}$ denotes the scalar curvature of spacetime metric $g_{\mu\nu}$ and $\omega$ is the coupling constant. Here we consider the case of $\omega=-3/2$. In this case, besides the spacetime diffeomorphism invariance, the action \eqref{Brans-Dicke Action} is also invariant under the following conformal transformations \cite{zhang2013loop}:
 \begin{equation}
 g_{\mu\nu}\to e^\lambda g_{\mu\nu},\ \phi\to e^{-\lambda}\phi .
 \end{equation} 

In the Hamiltonian formulation of the conformal invariant BD theory~\cite{olmo2011hamiltonian}, besides the Hamiltonian and diffeomorphism constraints, there exists also a conformal constraint that generates the above conformal transformations on the phase space.

 To introduce the elements of LQG for the quantum cosmology of this theory, we now recall the connection formalism of the conformal BD theory obtained in Ref.~\cite{zhang2013loop}. In this formalism,  the geometric canonical variables take the form of the gauge potential and electric fields from the $SU(2)$ Yang-Mills theory. Thus,  the canonical pair consists of an $SU(2)$ connection $A^i_a$ and a triad $E^a_i$ of density weight one, and they are related to the familiar triad variables in the following way. Denote the spatial triad fields as $e^a_i$ and the corresponding Levi-Civita spin connection as $\Gamma^i_a$. Let $\tilde{K}^i_b=\tilde{K}_{ab}e^{bi}$, $\tilde{K}_{ab}$ is defined in~\cite{zhang2011nonperturbative} and related to extrinsic curvature $K_{ab}$ of the spatial hypersurface through $\tilde{K}_{ab}=\phi K_{ab}$. The connection $A^i_a$ is related to  $\tilde{K}^i_a$ and $\Gamma ^i_a$ via $A^i_a=\Gamma^i_a+\gamma \tilde{K}^i_a$, where $\gamma$ is the Barbero-Immirzi parameter. Let $\pi_\phi$ be the momentum conjugate to $\phi$. Then, the  two canonically conjugate pairs satisfy the Poisson brackets
\begin{equation}
\begin{aligned}
\{A^i_a(x), E^b_j(y)\}&= \gamma \delta_a^b\delta_j^i\delta^3(x-y),\\
\{\phi(x),\pi_\phi(x)\}&=\delta(x,y).
\end{aligned}
 \end{equation}       

{As expected for any generally covariant canonical theory, the Hamiltonian appears as a linear combination of a set of first-class constraints \cite{zhang2011nonperturbative}. In our connection formulation, the constraint system consists of not only the familiar diffeomorphism and Hamiltonian constraints, but also the $SU(2)$-Gaussian and conformal constraints. In the Bianchi-I setting, the Gaussian and diffeomorphism constraints are trivially solved in the ``manifestly homogeneous gauge", in which the Hamiltonian and conformal constraints take the forms of \cite{zhang2011nonperturbative} }
\begin{equation}\begin{split}
\tilde{\mathcal{C}}_{H}=& \frac{\phi}{2\sqrt{q}}[F^j_{ab}-(\gamma ^2 +\frac{1}{\phi ^2})\epsilon^j_{mn} \tilde{K}^m_a \tilde{K}^n_b ]\epsilon_i^{jkl} E^a_k E ^b_l\\
&-\frac{3}{4 \phi} \sqrt{h}\left(D_{a} \phi\right) D^{a} \phi+\sqrt{h} D_{a} D^{a} \phi, \\
\mathcal{C}_{S}= &\tilde{K}^i_a E^a_i-\pi_\phi \phi ,
\end{split}\end{equation}
where $F_{ab}^i\equiv 2\partial_{[a}A^i_{b]}+\epsilon^i_{jk}A^j_aA^k_b$ is the curvature of the connection $A_a^i$.
The Poisson brackets between the two sets of constraints read \cite{zhang2011nonperturbative}
\begin{equation}\label{eq:poissonSH}
\{\mathcal{C}_{S}(\lambda),\tilde{\mathcal C}_H(M)\}=\tilde{\mathcal C}_H(\frac{\lambda M}{2}) ,
\end{equation}
in their smeared forms over the spatial hypersurface with $\lambda$ and $M$ denoting the smear functions.

We consider the spatially flat, homogenous Bianchi I model of cosmology~\cite{ashtekar2009loop,martin2009further}. Our spatial manifold $\Sigma$ will be topological $\mathbb{R}^3$. As a standard treatment, we restrict ourselves to diagonal Bianchi I metrics. Then the line element of the spacetime metric reads 
 \begin{equation}\label{line element}
 ds^2=-\tilde{N}^2dt^2+a_1^2(t) dx_1^2+a_2^2(t) dx_2^2+a_3^2(t) dx_3^2
 \end{equation}
where $\tilde{N}$ is the lapse function and $a_i$  are the scale factors. To avoid the non-compact problem of $\Sigma$, we introduce an elementary cell $\mathcal{V}$ and restrict all integrations to it. Fixing a fiducial metric $^oq_{ab}$ on $\Sigma$ with line element
\begin{equation}
 ds^2_o=dx_1^2+ dx_2^2+dx_3^2.
 \end{equation}
We denote by $L_i$ the lengths of the three edges of $\mathcal{V}$ measured by $^oq_{ab}$, then the volume measured by the fiducial metric is $V_o=L_1L_2L_3$. Also we introduce the fiducial cotriads $^ow^i_a\equiv D_a x^i$ and its dual fiducial triads $^oe^a_i$, satisfying $^oq_{ab}={^ow^i_{a}}{^ow^j_{b}}\delta_{ij}$ , where the Einstein's summation convention has been used. The physical cotriads are irrelevant to the fiducial rescales, and thus given by $w^i_a=a^i{^ow^i_a}$ (no summation is taken for $i$ here). The physical 3-metric $q_{ab}$ is given by $q_{ab}=w^i_a w^j_b\delta_{ij}$.

With those fiducial structures at hand, we can now introduce the symmetry-reduced connection variables\cite{ashtekar2009loop}. Under the Bianchi I symmetry, from each gauge equivalence class of these pairs we can select one and only one, given by
\begin{equation}
 A^i_a=:c^i(L^i)^{-1} \,^ow^i_a,  \text{and}
 \end{equation}
 \begin{equation}
 E^a_i\equiv \sqrt{q} e^a_i=:p_iL_i V^{-1}_o \sqrt{^oq} \,^oe^a_i,
 \end{equation}
where ${^oq}$ is the determinant of the fiducial metric and  $q=(| p_1p_2p_3 |)^oq V^{-2}_o$ is the determinant of the physical spatial metric $q_{ab}$. The momentum $p_i$ are directly related to the directional scale factor by
\begin{align}
 &p_1=\sgn(a_1)|a_2 a_3| L_2 L_3,      \nonumber\\
   &   p_2=\sgn(a_2)|a_1 a_3| L_1 L_3,  \nonumber\\
  & p_3=\sgn(a_3)|a_1 a_2| L_1 L_2,
 \end{align}
where we take the directional scale factor $a_i$ to be positive if the triad vector $e^a_i$ is parallel to ${^o}e^a_i$ and negative if it is antiparallel.

As the scalar field $\phi$ is spatially homogenous, its momentum reads $p_{\phi}=\int_{\mathcal{V}}\pi_\phi$.  The Hamiltonian of the model is given by $H(N,\lambda)= N C_{H} + \lambda C_{S}$ since the Gauss and diffeomorphism constraints vanish trivially. Here the multiplier $N$ is related to the lapse function $\tilde{N}$ in expression~\eqref{line element} by
\begin{equation}
\label{lapse function}
N= \frac{1}{\phi\sqrt{ |p_1p_2p_3 |}}\, \tilde{N}. 
\end{equation}
Then, the rescaled scalar and conformal constraints in the reduced variables are given by 
\begin{equation}
\begin{split}
\label{original constraints}
C_H&=-\frac{1}{\gamma^2}(c_1c_2p_1p_2+c_1c_3p_1p_3+c_2c_3p_2p_3),\\
C_{S} &=\frac{1 }{\gamma}(c_1p_1+c_2p_2+c_3p_3)-\phi  p_{\phi},\\
\end{split}
\end{equation}
in terms of the canonical coordinates satisfying the Poisson brackets
\begin{equation}\begin{split}
 \{c^i,p_i \}=\gamma \delta^i_j\, \text{and} \,\,
 \{\phi,p_{\phi}\}=1.
 \end{split}\end{equation}

It should be noted that the conformal constraint $C_S(\lambda)$ would generate constant rescalings of the spacetime metric rather than conformal transformations in the spatially homogeneous model due to the constant multiplier $\lambda$. However, it is easy to check that the Poisson bracket between the conformal constraint and the original Hamiltonian constraint in the Bianchi I model takes the same form as in  {\eqref{eq:poissonSH}}. Since the sign of $\phi$ is preserved on every trajectory generated by ${C}_{H}$ and ${C}_{S}$, we will restrict our discussion to the sector with the positive $\phi$ values. In this sector, we may choose the new canonical coordinates 
\begin{equation}
\{\tilde{\phi},P_{\tilde{\phi}}\}:=\{ \ln \phi,\phi p_{\phi}\}=1 ,
\end{equation}
with which the conformal constraint is further simplified to 
\begin{eqnarray}
\label{constraints3}
C_{S}=\frac{1}{\gamma}(c_1p_1+c_2p_2+c_3p_3)- P_{\tilde\phi} .
\end{eqnarray}
The $\phi>0$ sector of the Bianchi I  theory, governed by the above constraints, is what we will proceed to quantize using the techniques of LQG.

\section{Timeless conformal-BD LQC model}

\subsection{Quantum kinematics}

{ For the loop quantization of the cosmological model,  we follow the foundation provided in the series of works \cite{Metsaev:2006zy,martin2009further,martin2009physical,martin2008loop} for the quantum kinematic Hilbert space. The full kinematic Hilbert space consists of the unconstrained ``scalar" and ``tensor" sectors in the homogenous Bianchi-I setting. The scalar sector is quantized by the standard Schr\"odinger representation, with the quantum states of the square-integrable wave functions $L^2(\mathbb{R},d\tilde{\phi})$ over $\tilde{\phi}$, and the operators $\hat{\tilde{\phi}}$ and $$\hat{P}_{\tilde{\phi}}\equiv -i \hbar {\partial}_{\tilde{\phi}}$$ acting respectively as multiplication and differentiation operators. The tensor sector is quantized using the loop quantization procedure, leading to the basis of the ``quanta of the discretized spatial metric". We will construct our kinematic Hilbert space from a generic superselected sector preserved by the quantum dynamics. 
As it turns out, this space can be labeled by chosen numbers $\epsilon_i\in (0, 2)$ and spanned by a particularly chosen eigenbasis of triad operators $\{\hat{p}_i \}_{i=1,2,3}$, with the eigenvalues of a lattice based on the chosen $\epsilon_i$. This basis is given by 
 \begin{equation}
\begin{split}
\label{basis}
 \{\,\ket{v_1,v_2,v_3}\,|\, \,v_i=\epsilon_i +2 z_i \,; \,z_i\in \mathbb{Z}\,\}
\end{split}
\end{equation}
with
 \begin{equation}
\begin{split}
\hat{p}_i\ket{v_1,v_2, v_3}=&\sgn(v_i)\xi\gamma\ell_{Pl}^2|v_i|^{2/3}\ket{v_j}\,;\,\xi=(\frac{9 \sqrt{3} \pi }{4 \gamma })^{1/3}.
\end{split}
\end{equation}
The operator $\widehat{V}$ representing the metric-volume of the spatial cell $\mathcal V$ is given by 
\begin{eqnarray}
\widehat{V}=\big( \widehat{V}_1\widehat{V}_2\widehat{V}_3\big)^{\frac{1}{3}}\,; \,\widehat{V}_i\equiv |\hat{p}_i|^{\frac{3}{2}}.
\end{eqnarray}
Note that the specified interval $\epsilon _i\in (0, 2)$ corresponds to all the possible eigrenvalue-lattice of ${v}_i$ -- except for the single case with the lattice containing the point ${v}_i=0$. When the spectrum of $\widehat{V}$ contains zero, the inverse volume factors in the constraints require special quantization corrections to avoid the divergences. While these ``inverse-volume corrections" are interesting and well-studied \cite{ashtekar2004background,han2007fundamental,ashtekar2003mathematical}, we will
 avoid them in this work for technical simplicity by chosing $\epsilon _i\in (0, 2)$ instead of $\epsilon _i\in [0, 2)$.

{
Different schemes for the dynamical prescription in the Bianchi-I model of LQC \cite{ashtekar2003mathematical,ashtekar2011loop,ashtekar2009loop,McNamara:2022dmf,Metsaev:2006zy,martin2008loop} have been proposed. In the loop representation, the curvature variables $c_i$ appear only through the ``parallel transports" they generate-- via the holonomies over the paths in the three directions with the (coordinate-) lengths given by certain $\bar\mu_i$. This is the main feature of LQG that is captured in the LQC models. 
 However, a significant distinction between the two best-known schemes lies in the triad dependence of the $\bar\mu_i$ in the construction of the scalar constraint. In this work, we adopt the scheme proposed in \cite{Metsaev:2006zy,martin2008loop} where the holonomy operators in the scalar constraint are defined involving the curvature variables as
 \begin{equation}
\begin{split}
\label{parameters}
\hat{\mathcal{N}}_{\bar{\mu}_i}\equiv \widehat{\exp(i\bar\mu_i c_i)} \,; \,\bar{\mu}_i=\sqrt{\Box/|p_i|}\,;\,\Box=\frac{16}{9}\xi^3 \gamma\ell_{Pl}^2 .
\end{split}
\end{equation}
Their action on the basis reads
 \begin{equation}
\begin{split}
\hat{\mathcal{N}}_{\bar{\mu}_1}\ket{v_1,v_2, v_3}=\ket{v_1+2 ,v_2, v_3},\\
\hat{\mathcal{N}}_{\bar{\mu}_1}^\dagger\ket{v_1,v_2, v_3}=\ket{v_1-2,v_2, v_3}.
\end{split}
\end{equation}
It has been argued \cite{Chiou:2007mg} that, the LQC scalar constraint resulting in this way may be more faithful to the standard construction of the scalar constraint operator of the full LQG, as well as to the full theory’s notion of discrete quantum geometry. The other scheme proposed in \cite{ashtekar2009loop,McNamara:2022dmf}
prescribes the $\bar\mu_i$ as depending on all of the triad variables $\{p_{1}, p_2, p_3\}$ in a more intricate manner. This scheme yields the numerical results \cite{ashtekar2009loop,McNamara:2022dmf} that may be more appealing in certain physical aspects, such as the independence of the semi-classical dynamics from the choice of the fiducial cell. 

As we will see, taking only the matrix elements of the quantum-constraint kernel projector as the input, our approach works for both of the above LQC schemes that may give such computable elements. Here, for the initial analytic study of the conformal LQC, we choose the first scheme for its simpler quantum constraints to which the analytic solutions can be well-studied. } In the most common implementation of this scheme, the crucial factors of $c_i p_i$ for our model's constraint system are quantized as $\hat{\Omega}_i$ in the ``holonomy-regularized" form
\begin{eqnarray}\label{fd}
&&\hat{\Omega}_i\equiv  \frac{-i}{4\sqrt{\Box}}|\hat{p}_i|^{3/4}\Big[(\widehat{\sin(\bar\mu_i c_i)} \widehat{\sgn}_i +\widehat{\sgn}_i \widehat{\sin(\bar\mu_i c_i)}\Big]|\hat{p}_i|^{3/4}\nonumber\\
&&\text{with} \,\,\,\widehat{\sin(\bar\mu_i c_i)}\equiv \hat{\mathcal{N}}_{\bar{\mu}_i}-\hat{\mathcal{N}}^{\dag}_{\bar{\mu}_i}\,\text{and}\,\,\,\widehat{\sgn}_i \equiv \sgn({\hat{v}_i}).
\end{eqnarray}
It has been shown that \cite{martin2009further,Metsaev:2006zy,martin2009physical}, acting in the space spanned by the basis \eqref{basis}, the operator $\hat{\Omega}_i$ is self-adjoint and further preserves both of the two subspaces with ${\sgn}_i=\pm1$, in each of which $\hat{\Omega}_i$ has a non-degenerate spectrum of the full real line. To account for this ``parity degeneracy", we introduce operator $\widehat{P\sgn}_i$ that acts as
 \begin{equation}
\begin{split}
\widehat{P\sgn}_j\ket{{\Omega}_i, {\sgn}_i}= \ket{{\Omega}_i, {\sgn}'_i}\,;\,{\sgn}'_{i}= (-1)^{\delta_{i,j}}{\sgn}_{i}\,.
\end{split}
\end{equation}
Here, the $\ket{{\Omega}_i, {\sgn}_i}$ denotes the common eigenstate of $\{\hat{\Omega}_i, \hat{\sgn}_i\}$ that lies in either of the $\pm$ sides of the $v_i$ lattice that is indicated by ${\sgn}_i$. Note that since $\hat{\Omega}_i$ has the non-degenerate spectrum of the real-line under each given ${\sgn}_i$,  we may also introduce its conjugate operator $$\hat{P}_{\Omega_i}\equiv -i \hbar {\partial}_{\Omega_i }$$ that also has a non-degenerate spectrum of the real-line under each given ${\sgn}_i$.

Combining the scalar and tensor sectors, the kinematic Hilbert space of our model is thus given by  
 \begin{align}
\mathbb{K}&=\mathrm{span}\big\{\ket{v_i,\tilde{\phi}}\big\}=\mathrm{span}\big\{\ket{\Omega_i,\sgn_i, \tilde{\phi}}\big\} \nonumber\\
&=\mathrm{span}\big\{\ket{P_{\Omega_i},P\sgn_i, P_{\tilde{\phi}}}\big\} \nonumber
\end{align}
where the symbol $\mathrm{span}\{ \ket{x}\}$ denotes the Cauchy completion of the finite-norm linear span of the basis $\{ \ket{x}\}$, with the normalization of the basis states provided as
 \begin{align}
\label{basis1}
& \langle v'_i,\tilde{\phi}'| v_i,\tilde{\phi}\rangle= \delta_{v'_i,v_i } \delta(\tilde{\phi}-\tilde{\phi}')\,, \nonumber \\
&\langle \Omega'_i,\sgn'_i, \tilde{\phi}' | \Omega_i,\sgn_i, \tilde{\phi}\rangle =\delta_{\sgn'_i,\sgn_i }\,\delta(\Omega_i-\Omega'_i)\,\delta(\tilde{\phi}-\tilde{\phi}'), \nonumber\\
&\langle P'_{\Omega_i},P\sgn'_i, P'_{\tilde{\phi}} | P_{\Omega_i},P\sgn_i, P_{\tilde{\phi}}\rangle=\delta_{P\sgn'_i,P\sgn_i }\,\delta(P'_{\Omega_i}-P_{\Omega_i})\nonumber\\
&\times\delta(P_{\tilde{\phi}}-P'_{\tilde{\phi}}),
\end{align}
where $\delta_{x,y}$ and $\delta(x-y)$ denote respectively the Kronecker-delta and Dirac-delta functions.

The wavefunctions over the infinite one-dimensional lattice $\{v_i\}$ for each $i$ can be transformed into the semidiscrete fourier representation \cite{hu2023application} with the basis $\{\ket{b_i}\}$ given by
\begin{align}
\label{b modes}
&\bigg\{\,\ket{b_i}\,;\,\,b_i\in [-\pi\hbar, \pi\hbar]\,\bigg\}\nonumber\\
\equiv& \left\{\,\frac{1}{\sqrt{2\pi\hbar}}\sum_{v_i=\epsilon_i+2 z_i} e^{-i[\,(b_iv_i/2\hbar)+\eta(b_i)]}\ket{v_i}\,\right\}, 
\end{align}
where the ambiguous phase $\eta(b_i)$ is assigned as an arbitrary function of $b_i$. From \eqref{basis1} and \eqref{b modes}, it is straight forward to show that the states $\{\ket{b_1,b_2,b_3}\}$ are orthogonal with the Dirac-delta normalization, and the normalization factor ${1}/\sqrt{2\pi\hbar}$ can be understood using the ``standard phase" of $\eta(b_i)= -b_i\epsilon_i/2\hbar$ such that the wavefunctions become explicitly periodic in $b_i$ over $2\pi\hbar$. Therefore, in this new basis we have 
\begin{equation} 
\begin{split}
\mathbb{K}=\mathrm{span}\big\{\ket{b_i,\tilde{\phi}}\big\} \,;\,  \langle b'_i,\tilde{\phi}' |b_i,\tilde{\phi}\rangle= \delta(b_i-b'_i)\delta(\tilde{\phi}-\tilde{\phi}').
\end{split}
\end{equation}
Moreover, we may also introduce the operator $\hat{b}_i$ defined by its action of
$$\hat{b}_j \ket{b'_i}\equiv b'_j\ket{b'_i},$$ and the holonomy operators can then be expressed as
$$\hat{\mathcal{N}}_{\bar{\mu}_{i}}= e^{-2i\hat{b}_i/\hbar}.$$

\subsection{Physical Hilbert space and rigging map}

Based on the above quantum kinematics, we now specify a natural scheme for constructing the quantum constraint system generating the symmetries of the model in the Planck scales. As mentioned, we first introduce a specific form of regularization to the original constraints in \eqref{original constraints} using the loop variables $\{p_i,\sin(\frac{2b_i}{\hbar}), \cos(\frac{2b_i}{\hbar}) \}$, such that the quantum geometric effect can be introduced. Here we choose the simplest regularized forms as given by 
\begin{equation}
\begin{split}
\label{regularized constraints}
{C}^L_{H}=-\frac{1}{ \gamma^2}({\Omega}_1{\Omega}_2+{\Omega}_2{\Omega}_3+{\Omega}_3{\Omega}_1) , \\
{C}^L_{S}=\frac{1}{\gamma}({\Omega}_1+{\Omega}_2+{\Omega}_3)-{P}_{\tilde{\phi}}.\\
\end{split}
\end{equation}
Then, using the kinematic quantum operators representing the loop variables, we may directly canonically quantize the regularized constraints as
\begin{equation}
\begin{split}
\label{lqc constraints}
\hat{C}_{H}=-\frac{1}{ \gamma^2}(\hat{\Omega}_1\hat{\Omega}_2+\hat{\Omega}_2\hat{\Omega}_3+\hat{\Omega}_3\hat{\Omega}_1) , \\
\hat{C}_{S}=\frac{1}{\gamma}(\hat{\Omega}_1+\hat{\Omega}_2+\hat{\Omega}_3)-\hat{P}_{\tilde{\phi}}.\\
\end{split}
\end{equation}
It is straightforward to verify that $\hat{C}_H$ and $\hat{C}_S$ commutes, while the commutator between $\hat{C}_S$ and the original Hamiltonian constraint operator still mimics their Poisson bracket \eqref{eq:poissonSH}. In this sense, the obtained quantum constraint system is anomaly-free.
 
As in full LQG, the rigging map operator $$\hat{\mathbb{P}}:\mathcal S \to\mathcal S^*$$ is proposed as a generalized kernel projector for the quantum constraints\cite{thiemann2007modern}. Here $\mathcal S\subset \mathbb{K}$ denotes some dense subspace of elements analogous to the test functions in quantum mechanics, and $\mathcal S^*$ is the topological dual of $\mathcal S$, an analogue of the space of tempered distributions. It is expected that the image of $\hat{\mathbb{P}}$ can serve as the physical Hilbert space $\mathbb H\subset \mathcal{S}^*$, with the inner product between two physical states naturally given by the associated matrix element of $\hat{\mathbb{P}}$. In our model, the rigging map is defined as
\begin{equation}\begin{split}
\label{rigging map}
\hat{\mathbb{P}}\equiv&\delta(\hat{C}_S)\delta(\hat{C}_H)\\
:=&  \gamma^2\sum_{\sgn_i}\int \frac{d\Omega_2d\Omega_3}{|\Omega_2+\Omega_3|}\\
&\times|\Omega_2,\Omega_3,\underline{\Omega}_1,\sgn_i,\underline{P}_{\tilde{\phi}}\rangle\langle\Omega_2,\Omega_3,\underline{\Omega}_1,\sgn_i,\underline{P}_{\tilde{\phi}}|,
\end{split}\end{equation}
where $\underline{\Omega}_1(\Omega_2,\Omega_3)$ and $\underline{P}_{\tilde{\phi}}(\Omega_2,\Omega_3)$ denote the unique solution to the regularized constraints Eq.~\eqref{regularized constraints}, and they are given by
\begin{equation}\begin{split}\label{solution}
\underline{\Omega}_1(\Omega_2,\Omega_3)&:=-\frac{\Omega_2\Omega_3}{\Omega_2+\Omega_3},\\
\underline{P}_{\tilde{\phi}}(\Omega_2,\Omega_3)&:=\frac{1}{\gamma}(\Omega_2+\Omega_3-\frac{\Omega_2\Omega_3}{\Omega_2+\Omega_3}).
\end{split}
\end{equation}
Note that since the two quantum constraints commute, the ordering of delta functions in the rigging map is irrelevant. By construction, each physical state $\Psi$ is the image of some kinematic state $\psi$  under the rigging map $\hat{\mathbb{P}}$ and takes the form
\begin{equation}
\label{physwf}
\begin{split}
|\Psi)=:\hat{\mathbb{P}}|\psi\bigl>=&\delta(\hat{C}_S)\delta(\hat{C}_H)|\psi\rangle,
\end{split}
\end{equation}
with which the physical inner product can be naturally defined by
\begin{equation}
(\Psi'|\Psi):=\langle\psi'|\hat{\mathbb{P}}|\psi\rangle.
\end{equation}

Let us study the matrix elements of the rigging map $\hat{\mathbb{P}}$ containing the full physical information of the model, using the explicit form of the map provided in \eqref{rigging map}. The values of this rigging map's matrix elements in basis such as  $\{\ket{v_i,\tilde{\phi}}\}$ or $\{\ket{b_i,\tilde{\phi}}\}$ contain factors $\langle \tilde{\phi} | P_{\tilde{\phi}}\rangle$,$\langle{v_i|\Omega_i,\sgn_i }\rangle$ or $\langle b_i|\Omega_i, \sgn_i\rangle$. They are all readily available by the existing analytic calculation, via
\begin{equation}
\begin{aligned}
\left\langle\tilde\phi | P_{\tilde\phi}\right\rangle &= e^{i P_{\tilde\phi} \tilde\phi / \hbar},
 \\\langle b_i| \Omega_i, \sgn_i\rangle &=\frac{1}{\sqrt{2 \pi \hbar}} \sum_{v_i} {e^{i[\,(b_iv_i/2\hbar)+\eta(b_i)]}}\langle v_i | \Omega_i, \sgn_i\rangle.
 \end{aligned}
\end{equation}
The analytic form of the inner product $\langle{v_i|\Omega}_i, \sgn_i\rangle$ is also given in \cite{martin2009further}, and here we will simply study the lowest order terms in $\hbar$ and $\ell_{\text{Pl}}$ to understand the results.  
First, according to Eq.~\eqref{fd} the $\sgn_i$-preserving operator $\hat{\Omega}_i$ when restricted to each $\sgn_i=\pm1$ sector becomes
 \begin{equation}
 {\hat{\Omega}_i|_{\sgn_i=\pm1}=\pm\frac{1}{2\sqrt{\Box}}\sqrt{|\hat{V}_i|}\,\widehat{\sin(\frac{2b_i}{\hbar})}\,\sqrt{|\hat{V}_i}|}.
 \end{equation}
It can be verified using this form that the common eigenstates of $\hat{\Omega}_i$ and $\widehat{\sgn}_i$ can be expressed as \cite{lewandowski2018quantum}
{
 \begin{align}
 |\Omega_i, \pm1\rangle &=\sqrt{|\hat{v}_i|}\,\Theta(\pm\hat{v}_i)\nonumber
\\
&\times \int \frac{db_i}{\sqrt{|\Omega}_i|}\exp \mp i\left(\,\frac{\sqrt{\Box}\Omega_i }{(\xi\gamma)^{3/2}\ell^3_{\text{Pl}}}\ln\bigl |\tan(\frac{b_i}{\hbar})\bigl | \,\right)\,|b_i\rangle \nonumber\\
 &=:\sqrt{|\hat{v}_i}|\,|\ast\Omega_i,\sgn_i\rangle.
 \end{align}
 }
Then, it is straightforward to check that the values of $\bigl<v_i|\Omega_i,\sgn_i\bigl>=\sqrt{v_i}\bigl<v_i |\ast\Omega_i,\sgn_i\bigl>$ are given by
\begin{align}
\bigl<\Omega_i,\sgn_i|v_i\bigl>=&\sqrt{|v_i|}\int_{I_+}db_i\,\big<\ast\Omega_i,\sgn_i|b_i\bigl>\bigl<b_i|v_i\bigl>  \nonumber\\
&+\sqrt{|v_i|}\int_{I_-}db_i \big<\ast\Omega_i,\sgn_i|b_i\bigl>\bigl<b_i|v_i\bigl>,
\end{align}
where {$I_{\pm}=\{b_i|\pm\cos(\frac{2b_i}{\hbar})>0\}$}. The complex-phase factors of the integrands above are given by
\begin{align}
\label{stationary phase}
\theta(b_i,\Omega_i, V_i) &\equiv \sgn_i \,i\ln\bigg(\frac{\big<\ast\Omega_i,\sgn_i|b_i\bigl>\bigl<b_i|v_i\bigl>}{|\big<\ast\Omega_i,\sgn_i|b_i\bigl>\bigl<b_i|v_i\bigl>|}\bigg) \nonumber\\
&=-\frac{1}{(\xi\gamma)^{3/2}\ell^3_{\text{Pl}}}\bigg(\sqrt{\Box}\Omega_i\ln\bigl|\tan(\frac{b_i}{\hbar})\bigl|-\frac{b_iV_i}{\hbar}\bigg). 
\end{align}
 Here we note that for each pair $(\Omega_1, V_1\in I_{\Omega_1})$ within the semiclassical interval
$$I_{\Omega_1}\equiv\bigg[\,\,\sqrt{\Box}\,|\Omega_1|\,\,,\,\, \infty\,\,\bigg), $$ there is always one pair of solutions $\{b_1^+(\Omega_1,V_1)\in I_+,b_1^-(\Omega_1,V_1)\in I_-\}$ for $b_1$ giving the stationary phase via satisfying
\begin{equation}
\frac{\partial}{\partial b_1}\theta(b_1,\Omega_1,V_1)\bigl |_{b_1=b_{\pm}(\Omega_1,V_1)}\equiv 0.
\end{equation}
These solutions $b_1^{\pm}$ can be checked to be given by the expected semiclassical relation
\begin{equation}
{\frac{V_1}{2 \sqrt{\Box}} \bigl |\sin\frac{2b_1^{\pm}(\Omega_1,V_1)}{\hbar}\bigl|=\bigl|\Omega_1\bigl|}.
\end{equation}
Since $\partial_{b_1} \theta_1$ is of the order of $O(\ell^{-3}_{\text{Pl}}\hbar^{-1})=O(\ell^{-5}_{\text{Pl}})\gg1$,  we can employ the stationary phase approximation and obtain
\begin{align} 
 &\int_{I_{ \pm}} d b_i \langle\Omega_i, \sgn_i | b_i\rangle\langle b_i | V_i, \sgn_i  \rangle \,|_{V_i\in I_{\Omega_i}} \nonumber \\
 &=B^{ \pm}(\Omega_i, V_i) e^{-i \theta^{ \pm}(\Omega_i, V_i)} +O\left(\ell_{\text{Pl}}^{\,5/2}\right) ,
\end{align}
where we have
\begin{equation}
\begin{aligned}
\label{thetapm}
\theta_{ \pm}(\Omega_i, V_i) & \equiv \theta\left(b_i^\pm{(\Omega_i, V_i)}, \Omega_i, V_i\right) \pm \frac{\pi}{4},\,\, \text { and } \\ 
B_{ \pm}(\Omega_i, V_i) & \equiv \sqrt{\frac{2 \pi |v_i| }{|\Omega_i|\,|\partial_{b_i}^{2} \theta\left(b_i^\pm(\Omega_i, V_i), \Omega_i, V_i\right)|}}.
\end{aligned}
\end{equation}
Written out explicitly, the above says 
\begin{equation}
\begin{aligned} 
\label{asym}
&\int_{I_{ \pm}} d b_i\langle\Omega_i,\sgn_i | b_i\rangle\langle b_i | V_i,\sgn_i\rangle\,|_{V_i\in I_{\Omega_i}}\nonumber\\
&= \sqrt{\frac{3 \pi \hbar v_i}{2 \gamma^{1/2} \Omega_i^{2}\big|\partial^{2}_{b_i} \ln | \tan (b_i^{ \pm} / \hbar)| \big|}} \\ & \times \exp \,\frac{ -i\,\sgn_i\,}{(\xi\gamma)^{3/2}\ell^3_{\text{Pl}}}\bigg(\sqrt{\Box}\Omega_i\ln\bigl|\tan(\frac{b_i}{\hbar})\bigl|-\frac{b_iV_i}{\hbar}\bigg)\nonumber\\
&+O\left(\ell_{\text{Pl}}^{\,5/2}\right) .
\end{aligned}
\end{equation}
The asymptotic expression of $\langle\Omega_i| V_i,\sgn_i\rangle$ is thus derived when $V_i\in I_{\Omega_i}$ . When $V_i$ decreases and falls outside of $I_{\Omega_i}$, the wave function $\langle\Omega_i| V_i,\sgn_i\rangle$ enters a purely quantum tunneling region. 

Recall that we have constructed the kinematic Hilbert space $\mathbb{K}$ as being spanned by the  basis of the lattice eigenvalues, with a generic starting value of $\epsilon_i$ that labels the superselected sector of our choice. Also, the wavefunction of the associated fourier basis state $\ket{b_i}$ contains an ambiguous phase given by the arbitrary $\eta(b_i)$. From the analysis in this section, it becomes clear that the values of $\epsilon_i$ and $\eta(b_i)$ have no effect on the obtained analytic forms of the key results, although they do control the values of certain involved parameters, such as those of the $V_i$ appearing from \eqref{stationary phase} to \eqref{thetapm}.

\section{Schr\"odinger theories of the model via augmented quantum reference frames}

Having constructed the physical Hilbert space of the model, we now proceed to explore its physical dynamics. 
At the full theory level, the canonical formulation of the linking theory ( conformal BD theory ) is purely constrained by the first-class constraints $\{ \mathcal{C}_H, \mathcal{C}_{D_k}, \mathcal{C}_S\}$ defined in the kinematic phase space with the canonical coordinates of all the field conjugate pairs $\{(X_i, P_i)\}\equiv \{(\tilde\phi, P_{\tilde\phi}), (X_{\bar{i}}, P_{\bar{i}})\}$. Here the $\mathcal{C}_{D_k}$ with $k=1,2,3$ denotes the diffeomorphism constraints generating the spatial diffeomorphisms upon the fields $\{(X_i, P_i)\}$.

It has been shown that GR is recovered from the linking theory \cite{mercati2014shape} under the choice of imposing the gauge condition $\tilde\phi= 0$ while solving for its momentum $P_{\tilde\phi}$ with respect to the conformal constraints $\mathcal{C}_S$. This gives a symplectic reduction of the linking theory’s phase space into the ADM phase space of GR coordinatized by $\{(X_{\bar{i}}, P_{\bar{i}})\}$. Particularly, the remaining constraints $\{  \mathcal{C}_H, \mathcal{C}_{D_k}\}$ of the linking theory after the reduction becomes identical to the ADM constraints $\{  H, {D_k}\}$ of GR, consisting of the vector and scalar constraints. Thus, the problem of obtaining the physical dynamics in the linking theory under this gauge is identical to that in GR, which is the absence of an absolute notion of time due to the spacetime re-foliation invariance.

One of the prevailing approaches to solving the problem of time in GR is through the relational description, with the dynamics described in the physical spacetime coordinates given by a properly chosen set of reference fields in the system. In the canonical formulation, the physical states of spacetimes are equivalent to the set of the on-shell orbits of the ADM constraints $\{ H, {D_k}\}$.  Since the ADM constraints generate the Cauchy-hypersurface deformations, each point in one of the orbits represents a specific Cauchy hypersurface in the spacetime associated with the orbit. In the relational description, one may first choose a splitting of the canonical coordinates $\{(X_{\bar{i}}, P_{\bar{i}})\}$ into a reference sector $\{(X_{\bar\mu\,\text{GR}}, P_{\bar\mu\,\text{GR}})\}$ and dynamical sector $\{(X_{\bar{I}\,\text{GR}}, P_{\bar{I}\,\text{GR}})\}$. Then, one constructs the set of reference fields $T_{\bar\mu\,\text{GR}}(X_{\bar\nu\,\text{GR}}, P_{\bar\nu\,\text{GR}})$ using the reference sector and assign their instantaneous values as a set of specified functions $\{T_{\bar\mu\,\text{GR}}({t})\}$ over the time parameter ${t}$. For each moment of time ${t}$, the properly chosen reference fields and their instantaneous values should specify a unique point in every on-shell orbit. Physically, this means that the instantaneous values of the reference fields truly specify a well-defined reference frame for each of the spacetimes, such that each value in ${t}$ corresponds to a single hypersurface of each of the spacetimes, described in a given spatial coordinate system. For each value of $t$, the field values of the dynamical sector $\{(X_{\bar{I}\,\text{GR}}|_{T_{\bar\mu\,\text{GR}}({t})}, P_{\bar{I}\,\text{GR}}|_{T_{\bar\mu\,\text{GR}}({t})}\}$ taken at the specified point in each orbit are consequently specified. Using this fact, we may define a set of functions $\{(X_{\bar{I}\,\text{GR}}({t}), P_{\bar{I}\,\text{GR}}({t})\}$ on the GR constraint surface, with their values constant in each orbit and given by $$\{(X_{\bar{I}\,\text{GR}}({t})\equiv X_{\bar{I}\,\text{GR}}|_{T_{\bar\mu\,\text{GR}}({t})}\,, \,\,P_{\bar{I}\,\text{GR}}({t})\equiv P_{\bar{I}\,\text{GR}}|_{T_{\bar\mu\,\text{GR}}({t})}\}.$$ While capturing the local values of the dynamical fields in the reference frame specified via the reference fields, these invariant functions serve as the relational observables depending only on the physical states.

Note that in the above procedure of obtaining the physical dynamics of GR from the linking theory,  the treatment of the full constraint system $\{ \mathcal{C}_H, \mathcal{C}_{D_k}, \mathcal{C}_S\}$ seems to combine of two drastically different methods. That is, while the symplectic reduction with respect to $\mathcal{C}_S$ ``breaks" the conformal symmetry due to the gauge fixing, the relational observables $\{(X_{\bar{I}\,\text{GR}}({t}), P_{\bar{I}\,\text{GR}}({t})\}$ manifestly preserve the remaining symmetry. To explore the content of the linking theory as a unifying theory, one naturally seeks a fully gauge-invariant approach. This can be done, as it has been understood~\cite{lyakhovich2001extended,thiemann2006reduced} that the successful symplectic reduction procedures with respect to a first-class constraint system can always be reformulated into a relational prescription, using the generalized notion of relational observables associated to the constraint system.  

Therefore, we are motivated to apply the manifestly gauge-invariant treatment to the linking theory, using the generalized relational observables described below. In the canonical formulation of linking theory, the physical states are equivalent to the set of the on-shell orbits of the constraints $\{ \mathcal{C}_H, \mathcal{C}_{D_k}, \mathcal{C}_S\}$. From our previous discussion,  these physical states would contain those generated from applying arbitrary conformal transformations to the spacetimes.  Analogous to the relational description in GR, one may choose a splitting of the canonical coordinates $\{(X_i, P_i)\}$ into the reference sector $\{(X_\mu, P_\mu)\} \equiv \{(\tilde{\phi}, P_{\tilde{\phi}}), (X_{\bar\mu\,\text{GR}}, P_{\bar\mu\,\text{GR}})\}$ and the dynamical sector $\{(X_I, P_I)\} \equiv \{ (X_{\bar{I}\,\text{GR}}, P_{\bar{I}\,\text{GR}})\}$. Then, one constructs the set of reference fields $ \{ T_\mu(X_\nu, P_\nu)\}$ with the assigned instantaneous values through the functions $ \{ T_\mu(t)\}$.  For each moment of time $t$, the properly chosen reference fields and their instantaneous values should again specify exactly one point in each orbit of a proper sector of physical states. The field values of the dynamical sector $(X_{I}(t), P_{I}(t))\}$ taken at the specified point in each of the orbits then yield a set of functions of the physical states serving as the augmented relational observables, capturing the local values of the dynamical fields in an augmented notion of reference frame specified via the reference fields. Specifically, we may choose$ \{ T_\mu\}\equiv \{ \tilde{ \phi},\,T_{\bar{\mu}\,\text{GR}} \}$ with the assigned instantaneous values through the functions $ \{ T_\mu(t)\}\equiv\{ \tilde{\phi}(t)=0, T_{\bar\mu\,\text{GR}}(t)\}$. It then becomes clear that this prescription using the particular relational observables is indeed equivalent to the symplectic reduction procedure leading to GR, with the identification $ (X_{\bar{I}\,\text{GR}}(t)\, , \,P_{\bar{I}\,\text{GR}}(t))=(X_{I}(t)\,, \,P_{I}(t))$ and  $\{X_{\bar{I}\,\text{GR}}({t})\,, \,P_{\bar{I}\,\text{GR}}({t})\}_{\text{GR}}= \{X_{I}(t)\, , \,P_{I}(t))\}=1$. In this precise sense, GR is simply a representation of a specific sector of the linking theory through the chosen set of augmented relational observables.

The so-called SD can also emerge from the linking theory via a different symplectic reduction procedure. In this case, the full reduction procedure involves conditions upon $\{(\phi, P_{\phi}),(X_{\bar{\mu}\,\text{SD}}, P_{\bar{\mu}\,\text{SD}}),({S}, P_{S})\}$, where the pairs $\{X_{\bar{\mu}\,\text{SD}}, P_{\bar{\mu}\,\text{SD}}, S,  P_{S}\}$ are all specific functions of $\{(X_{\bar{i}}, P_{\bar{i}})\}$.  Here, $(X_{0\,\text{SD}}, P_{0\,\text{SD}})$ must be a single global degree of freedom for the spatial slice and also spatial-diffeomorphism invariant with $\{\mathcal{C}_{D_k}, X_{0\,\text{SD}}\}=\{\mathcal{C}_{D_k}, P_{0\,\text{SD}}\}=0$. In the existing literature \cite{ mercati2014shape,gielen2018gravity}, the prevailing choice for $X_{0\,\text{SD}}$ is given by the York time variable, with its conjugate momentum $P_{0\,\text{SD}}$ given by the total spatial volume. SD is a theory with an absolute notion of time, and its kinematic phase space for the moment $\tau$ can be obtained by the symplectic reduction procedure with respect to the BD-scalar constraints $\mathcal{C}_H$ under the assigned values of $(P_{\phi}(\tau)\,=0\,, \,\,X_{0\,\text{SD}}(\tau)\,)$. Specifically, the field condition $P_{\phi}(\tau)=0$ fixes the BD lapse functions $N$ up to a homogeneous mode, which is then determined by the global condition $X_{0\,\text{SD}}(\tau)$ specifying one unique slice for each $\tau$ in the given foliation. Along with solving the constraint $\mathcal{C}_H$, this reduction leads to the kinematic phase space of SD at the moment $\tau$. Also, the original constraints $\mathcal{C}_S$ and $\mathcal{C}_{D_k}$ are respectively reduced to the constraints $S$ and $D_k$, generating the conformal and spatial diffeomorphism gauge symmetries of SD. Finally, the physical dynamics of SD may be obtained by the symplectic reduction with respect to the constraints $S$ and $D_k$ respectively under the gauge-fixing conditions $ P_{S}=P_{S}(\tau)$ and $\{X_{\bar{\mu}\,\text{SD}}=X_{\bar{\mu}\,\text{SD} }(\tau)\}_{\bar{\mu}=1,2,3}$.

It is obvious that a manifestly gauge-invariant prescription of the SD may be given by a corresponding set of generalized relational observables in the linking theory. Clearly, this time we have the reference sector as $\{(X'_\mu, P'_\mu)\} \equiv \{(\phi, P_{\phi}),(X_{\bar{\mu}\,\text{SD}}, P_{\bar{\mu}\,\text{SD}}),({S}, P_{S})\}$, using $\{T'_\mu\} \equiv \{P_{\phi}, X_{\bar{\mu}\,\text{SD}}, P_{S} \}$ with the assigned values $\{T'_\mu(\tau)\}= \{P_{\phi}(\tau)=0, X_{\bar{\mu}\,\text{SD}}(\tau), P_{S}(\tau) \}$. Again, suppose we have the decomposition  $\{(X_i, P_i)\}= \{(X'_\mu, P'_\mu), (X'_I, P'_I) \}$, the physical dynamics of SD should be described by the augmented relational observables $\{X'_I(\tau), P'_I(\tau)\}$. As it is known, the physical states of GR and those of SD intersect as a non-trivial subset of physical states in the linking theory. GR and SD are thus equivalent to the two reduced phase-space representations in this sector of the linking theory. This equivalence becomes explicit using the relational observables since the subset of physical states is the common domain of the observables $\{X'_I(\tau), P'_I(\tau)\}$ and $\{X_{I}(t)\, , \, P_{I}(t))\}$, and {thus the precise transformation from the dynamics in GR into that in SD could be obtained from the calculable relation between  $(X'_I(\tau), P'_I(\tau))$ and $(X_{I}(t)\, , \, P_{I}(t)))$ in their common domain.}
 Actually, it is known that the set of SD physical states is a subset of GR physical states in the linking theory, consisting of the GR spacetimes allowing the maximal foliation with spatial slices of zero mean curvature $K=0$~\cite{ mercati2014shape,gielen2018gravity}.

In the previous sections, we have introduced the conformal BD LQC, as a model of quantum linking theory with the well-constructed physical Hilbert space. Now following the above motivation, we will obtain the Schr\"odinger theories of GR and SD cosmologies, using the associated augmented quantum relational observables in the physical Hilbert space. We will use the approach of quantum reference frames tailored for deriving a Schr\"odinger theory from a canonical quantum gravity theory governed by a constraint rigging map. Recently, an algorithm was formulated \cite{lewandowski2018quantum} for extracting Schr\"odinger theories under all viable physical notions of time, from the rigging map $\hat{\mathbb{P}}:\mathbb{K} \to \mathbb{H}$ solving a set of quantum constraints $\{\hat{C}_{\mu}\}$. Each of these Schr\"odinger theories refers to a certain set of quantum degrees of freedom from $\mathbb{K}$ as a reference sector, with their values specifying the moments of the physical time. The relevant elements of $\hat{\mathbb{P}}$, restricted to the specified subspaces for the reference sector, are then transformed by the algorithm into the unitary propagator for the corresponding relational Schr\"odinger theory for the dynamical sector. In the original setting of GR, the set $\{\hat{C}_{\mu}\}$ contains only the constraints associated with the spacetime diffeomorphism symmetry, and so the quantum reference frames embody the quantum general covariance. Thus all the derived Schr\"odinger theories emerge from one timeless canonical quantum theory governed by $\hat{\mathbb{P}}$, via the relational Dirac observables defined in $\mathbb{H}$. As we shall demonstrate, through its generalized application to our model here with $\{\hat{C}_{\mu}\}= \{\hat{C}_S , \hat{C}_H \}$, the approach of quantum reference frames embody the desired notion of the augmented reference frame at the quantum level, through the well-defined augmented quantum relational Dirac observables. Let us first provide an updated outline of the algorithm and its fundamental principle.

\subsection{Quantum reference frame algorithm under general setting}

We assume a Dirac theory of canonical quantum gravity governed by the quantum constraint $\hat{M}=\hat{M}^{\dagger}$ defined in the kinematic Hilbert space $\mathbb{K}=\mathrm{span}\{\ket{X_i}\}$  spanned by the eigenstates $|X_i\rangle$ of the complete set of self-adjoint operators. Then the constraint can be imposed by a rigging map operator 
$$\hat{\mathbb{P}}\equiv \delta(\hat{M}) \,:\,\mathcal{S} \to \mathcal{S}^*.$$
As mentioned earlier in the context of our model, the matrix elements of the rigging map naturally define the associated physical inner products in the physical Hilbert space $\mathbb{H}\subset \mathcal{S}^*$. Again, the inner product between two physical states $\{\,|\Psi_1)\equiv \hat{\mathbb{P}}\ket{\psi_1}, |\Psi_2) \equiv \hat{\mathbb{P}}\ket{\psi_2} \, \}\subset \mathbb{H}$ is defined by
\begin{eqnarray}
\label{inner product}
(\Psi_1|\Psi_2)\equiv \left\langle {{\psi _1}} \right|\mathbb{P}\left| {{\psi _2}} \right\rangle .
\end{eqnarray}
Note that the {dynamics has} to emerge from these physical states which are constructed without any notion of time. In our approach, we implicitly construct a complete set of local observables, called the elementary relational observables, identified using a quantum reference frame in which a specific set of reference quantum fields $\hat{T}_\mu$ takes a set of given values $T_\mu(t)$ at any moment $t$ of time. The calculation of the dynamics for these observables is given by an algorithm consisting of the following steps.

\begin{enumerate}
\item To specify a quantum reference frame, one first chooses from the set of basic self-adjoint operators in $\mathcal H_{\rm kin}$ a reference sector, with the rest being the dynamic sector. Concretely, we introduce a splitting to the total set of the conjugate pairs as $\{(X_i, P_i) \}= \{({X}_\mu, {P}_\mu)\}\cup \{({X}_I, {P}_I)\}$, where the two disjoint sets $\{({X}_\mu, {P}_\mu)\}$ and $\{({X}_I, {P}_I)\}$ represent respectively the reference and dynamic sectors. The specific reference quantum fields are constructed as $\{\hat{T}_\mu=\hat{T}_\mu(\hat{{X}}_\nu, \hat{{P}}_\nu)\}$, and the choice of a reference frame can be completed by specifying the set of functions $\{T_\mu(t)\}$ of the moments $t$, taking values from the spectrum of $\hat{T}_\mu$. It should be clear now that each $t$ is associated with a common eigenspace for the reference operators $\{\hat{T}_\mu\}$, given by  $$\mathbb{S}_t\equiv \ket{T_\mu(t)}\otimes\mathrm{span}\{\ket{{P}_I}\}\,\tilde{\subset} \,\mathcal{S}^*\,.$$ Here the expression $\mathbb{S}_t\tilde{\subset}\mathcal{S}^*$ denotes the relation of $\mathbb{S}_t$ being an generalized eigenspace of the operators $\{\hat{T}_\mu\}$ defined in $\mathcal{S}^*$ with potentially continuous spectra. Correspondingly, for each value of $t$ we assume that the $\hat{\mathbb{P}}$ is well-defined acting on $\mathbb{S}_t$, such that we have $$\hat{\mathbb{P}}|_{\mathbb{S}_t}:\,  \mathbb{S}_t\to \mathbb{D}_t  \,\tilde{\subset}\, \mathbb{H}\,,$$ with $\mathbb{D}_t$ being a generalized subspace of $\mathbb{H}$ in the same sense above.

\item One calculates the relevant transition amplitude matrix elements
\begin{eqnarray}
\label{amp1}
\hat{\mathbb{P}}_{t't}({X}'_I, {X}_I )\equiv \left\langle {{{\bar T}_\mu }(t'),{X}'_I} \right|\hat{\mathbb{P}} \left| {{{\bar T}_\mu }(t),{X}_I } ,\right\rangle
\end{eqnarray}
between arbitrary $t'$ and $t$. This matrix effectively defines an operator $\hat{\mathbb{P}}_{t't}: \mathbb{S}_t\to \mathbb{S}_{t'} $. Carrying the meaning of the inner products for the space $\mathbb{D}_t$, the square matrix $\mathbb{P}_{tt}$ in the dynamical sector must be diagonalizable into a diagonal matrix with non-negative real elements. Upon this given condition, the matrix $\mathbb{P}_{tt}$ for a valid quantum reference frame must further be densely positive-definite in $\mathbb{S}_t$, implying a bijection between $\mathbb{D}_t$ and $\mathbb{S}_t$ through the rigging map.

\item Suppose that the $\mathbb{P}_{tt}$ is indeed invertible. Then one may introduce the correction factor operator $\hat{\Lambda}_t \equiv \hat{\mathbb{P}}^{-\frac{1}{2}}_{tt}\,:\,\mathbb{S}_t\to\mathbb{S}_t $ which would be also (densely) well-defined and positive-definite. In matrix notation, the relational propagator $\hat{U}_{t't}\,:\,\mathbb{S}_t\to\mathbb{S}_{t'}$ is then defined as
\begin{eqnarray}
\label{inner product3}
\hat{U}_{t't}\equiv \,\hat{\Lambda}_{t'} \,\,\hat{\mathbb{P}}_{t't} \,\,\hat{\Lambda}_t\,.
\end{eqnarray}
Note that, when $t'=t$ by construction one has the desired condition $\hat{U}_{tt}= \hat{I}$. Upon this given condition, the square matrix $U_{t't}$ defined as 
\begin{eqnarray}
\label{U matrix}
{U}_{t't}({X}'_I, {X}_I)\equiv \left\langle {{{\bar T}_\mu }(t'),{X}'_I} \right|\hat{U}_{t't}\left| {{{\bar T}_\mu }(t),{X}_I } ,\right\rangle
\end{eqnarray}
for a valid quantum reference frame must further satisfy $U^{\dagger}_{t't}=U^{-1}_{t't}$, implying $\mathbb{D}_t= \mathbb{D}_{t'}\equiv\mathbb{D}$ and the time independent Heisenberg physical state space $\mathbb{D}$.

\item When the above holds, we derive a Schr\"odinger dynamics emerging from the timeless Dirac theory. The wave functions take the form $\Psi_{\mathbb{D}}[{X}_I](t)$ and is evolved by $\hat{U}_{t't}$.  The theory in its Heisenberg picture is given by the physical states in the domain $\mathbb{D}$, and the complete sets of elementary relational observables $\{\hat{{P}}_I(t)\,: \,\mathbb{D}\to \mathbb{D}\}$ and $\{\hat{{X}}_I(t)\,: \,\mathbb{D}\to \mathbb{D}\}$, which by construction enjoy their correspondent kinematic algebra
\begin{eqnarray}
\label{obsv alge}
 [\hat{{P}}_J(t),\hat{{X}}_I(t)] = [\hat{{P}}_J,\hat{{X}}_I](t) .
\end{eqnarray}

\end{enumerate}

The above implicitly defined elementary relational observables can be expressed directly in the form of 
   \begin{equation}
\label{rel obsv 0}
   (\hat {{X}}_I(t),\hat {{P}}_J(t)) \equiv \hat{\mathbb{P}}  \hat{\Lambda}_t \hat({\hat {{X}}_I},{\hat {{P}}_J}){\hat \Lambda_t  ^{ - 1}}{\hat{{\Pi}}_{t}}.
   \end{equation}
Here $\hat{\Pi}_{t}: \mathbb{D}\to \mathbb{S}_t$ denotes the inverse map of the bijection $\mathbb{S}_t\to \mathbb{D}$ under the rigging map. The eigenstates of Dirac observables take forms as
   \begin{equation}
\label{obsv eigenstate}
    \left| {\left. {{{X}}_I(t)} \right)} \right. \equiv \hat{\mathbb{P}} \hat \Lambda_t \left| {{{T}_\mu }(t),{{{X}}_I}} \right\rangle
   \end{equation}
  The wave function $\Psi_{\mathbb{D}}$ can be written as a function of elementary relational observables $\Psi_{\mathbb{D}}[{X}_I](t)$ or equally $\Psi_{\mathbb{D}}[{P}_I](t)$, and it is evolved by $\hat{U}_{t't}$ or equally the physical Hamiltonian.  Lastly, using~\eqref{obsv eigenstate} and \eqref{inner product} we can easily read off the form of  $\hat{\Pi}_{t}$ as
     \begin{equation}
\label{Pi op}
    \hat{\Pi}_{t}= \sum_{{X}_I} \hat{\Lambda}_t \ket{ {T}_\mu (t),{X}_I}\bra{ {T}_\mu (t),{X}_I}\hat{\Lambda}_t
   \end{equation} 

Finally, the physical Hamiltonian $\hat{\bm{H}}(t)$ of the Schr\"odinger theory can be given via its matrix elements as
\begin{equation}
\label{phys H}
({X}'_I(t)|\,\hat{\bm{H}}(t)\,|{X}_I(t))=i \hbar \,\partial_{t'}\, U_{t',t}({X}'_I;{X}_I)\, |_{t'=t}
\end{equation}

We will now use the relevant rigging map elements of our model to compute the unitary evolutions under the two augmented quantum reference frames, which have the most important significance in describing the linking theory.

\subsection{Frames with specified $(\tilde{\phi}(t),{V}^+_1(t))$ for GR LQC}

Here we choose  $$\{ (\hat{X}_{\mu}, \hat{P}_{\mu})\} \equiv\{(\hat{\tilde{\phi}},\hat{P}_{\tilde{\phi}}),(\hat{P}_{{\Omega}_1},\widehat{P\sgn}_1,\hat{\Omega}_1, \widehat{\sgn}_1)\}, $$leaving the dynamical sector as $$\{ (\hat{X}_{I}, \hat{P}_{I})\}\equiv \{(\hat{P}_{{\Omega}_I},\widehat{P\sgn}_I,\hat{\Omega}_I, \widehat{\sgn}_I)\}_{I=2,3}.$$ Just to be clear, here on we regard the $\{\hat{P}_{{\Omega}_I},\widehat{P\sgn}_I \}$ and $\{\hat{\Omega}_I, \widehat{\sgn}_I\}$ as the configuration and momentum operators. In the deformed classical theory governed by the classical constraints of the form \eqref{regularized constraints}, we would choose the background fields so that each given values of the dynamical variables $(v_2,b_2,v_3,b_3)$ together with a given proper background values $(T_1,T_2)$ correspond to one unique point on the constraint surface. We will study one particular choice satisfying this requirement, given by $\{{T}_1, {T}_2\}\equiv\{{\tilde{\phi}},{V}^+_1\}$ with $V^+_1\equiv \Theta({\sgn}_1)\Theta (\cos(\frac{2b_1}{\hbar}))\,v_1$ where $\Theta(x)=1$ when $x>0$, and $\Theta(x)=0$ otherwise. Correspondingly, in the quantum theory we simply choose the background quantum fields to be $$\{\hat{T}_1, \hat{T}_2\}\equiv\{\hat{\tilde{\phi}}\,,\,\hat{V}^+_1\}$$ with
\begin{equation}
\ket{{V}^+_1}\equiv \Theta(\widehat{\sgn}_1)\Theta{(\widehat{\cos(\frac{2b_1}{\hbar})})}\ket{V_1,{\sgn}_1}\bigg|_{V_1=V^+_1,\,{\sgn}_1=+1 }.
\end{equation}
We then specify the appropriate functions $\{\tilde{\phi}(t),V^+_1(t)\}$ as a family of the eigenvalues parametrized by the continuous time parameter $t$. Thus for each value of $t$ we have the associated eigenspace $${\mathbb{S}^{t}=\ket{\tilde{\phi}(t),V^+_1(t)}\otimes \mathrm{span}\{|\Omega_I,{\sgn}_I \bigl>\}.} $$

\subsubsection{  Schr\"odinger theory under arbitrary $\tilde{\phi}(t)$ assignment}

By using the rigging map \eqref{rigging map}, the relevant transition amplitudes $(\Omega_I,{\sgn}_I, t)\rightarrow(\Omega'_I, {\sgn}'_I,t')$ can be computed as
\begin{align}
&\mathbb{P}_{t',t}(\Omega'_I,{\sgn}'_I;\Omega_I,{\sgn}_I) \nonumber\\
=&\langle\tilde{\phi}(t'),V^+_1(t'),\Omega'_I,{\sgn}'_I|\hat{\mathbb{P}}|\tilde{\phi}(t),V^+_1(t),\Omega_I,{\sgn}_I\rangle \nonumber\\
=& \gamma^2 \frac{ \langle V^+_1(t')|\underline{\Omega}_1\rangle\langle\underline{\Omega}_1|V^+_1(t)\rangle}{|\Omega_2+\Omega_3|}e^{i\underline{P}_{\tilde{\phi}}(\tilde{\phi}(t')-\tilde{\phi}(t))/\hbar} \nonumber\\
&\times\delta_{{\sgn}'_I,{\sgn}_I}\delta(\Omega'_2-\Omega_2)\delta(\Omega'_3-\Omega_3).
\end{align}
Taking the next step of the quantum reference frame algorithm, we then find
\begin{equation}\label{lambda}
\hat{\Lambda}_t=\mathbb{P}_{t,t}^{-1/2}=\frac{ \sqrt{|\hat{\Omega}_2+\hat{\Omega}_3|} }{\gamma|\widehat{\bigl<V^+_1|\underline{\Omega}_1\bigl>}|},
\end{equation}
 where we defined $$\widehat{\bigl<V^+_1|\underline{\Omega}_1\bigl>}\ket{\Omega_I,{\sgn}_I}\equiv\bigl<V^+_1|\underline{\Omega}_1(\Omega_I)\bigl>\,\,\ket{\Omega_I,{\sgn}_I}.$$ Since the operator $\hat{\Lambda}_t$ is clearly densely defined in $\mathbb{S}_t$, we know that the restricted rigging map $\hat{\mathbb{P}}|_{\mathbb{S}_t} :\mathbb{S}_t\to \mathbb{D}_t$ is indeed non-degenerate for every $t$.

Next, we examine if the condition $\mathbb{D}_t=\mathbb{D}_{t'}$ is true as desired by computing
\begin{equation}\begin{split}
\label{propagator}
&U_{t',t}(\Omega'_I, {\sgn}'_I;\Omega_I, {\sgn}_I)\nonumber\\
&=(\Lambda^{\dag}_{t'}\mathbb{P}_{t',t}\Lambda_t)(\Omega'_I, {\sgn}'_I;\Omega_I, {\sgn}_I)\nonumber\\
&=\delta_{{\sgn}'_I,{\sgn}_I}\delta(\Omega'_2-\Omega_2)\delta(\Omega'_3-\Omega_3)\nonumber\\
&\,\,\,\,\times e^{i\,\underline{P}_{\tilde{\phi}}(\tilde{\phi}(t')-\tilde{\phi}(t))/\hbar}e^{i({\theta}_1(\Omega_2,\Omega_3,\,t')-{\theta}_1(\Omega_2,\Omega_3,\,t))},
\end{split}\end{equation}
where we introduce
\begin{eqnarray}
 e^{-i\theta_1(\Omega_2,\Omega_3,\,t)}\equiv\frac{\bigl<\underline{\Omega}_1(\Omega_2,\Omega_3), +1|V^+_1(t)\bigl>}{|\bigl<\underline{\Omega}_1(\Omega_2,\Omega_3), +1|V^+_1(t)\bigl>|}.
\end{eqnarray}
One can see now that $U_{t',t}$ is indeed given by a unitary matrix, therefore we have $\mathbb{D}_{t}=\mathbb{D}_{t^{'}>t} \equiv \mathbb{D}_{\text{ST}}$. So the proposed $t$ serves as a physical time for a Schr\"odinger theory of the linking theory.

Finally, for each assignment of the functions $\{\tilde{\phi}(t),V^+_1(t)\}$ we have derived a Schr\"odinger theory from the Dirac theory. In the Heisenberg representation, the space of physical states is $\mathbb{D}_{\text{ST}}\subset \mathbb{H}$, and the complete set of elementary quantum relational observables is given by $$\{ (\hat{X}_{I}(t), \hat{P}_{I}(t))\}\equiv \{(\hat{P}_{{\Omega}_I}(t),\widehat{P\sgn}_I(t),\hat{\Omega}_I(t), \widehat{\sgn}_I(t))\}\,.$$ Note that the above is derived for the reference field values $\{\tilde{\phi}(t), V^+_1(t)\}$ at each $t$ arbitrarily assigned, and the corresponding Heisenberg state space ${\mathbb{D}_{\text{ST}}}$ is thus the same for all such possible assignments. The associated family of Schr\"odinger theories thus represent the same Dirac theory in ${\mathbb{D}_{\text{ST}}}$ as a realization of the  ``multi-finger" Schr\"odinger dynamics.

To further understand the derived Schr\"odinger dynamics, we extract the physical Hamiltonian $\hat{\bm H}(t)$ through \eqref{phys H} by computing the matrix $U_{t't}$. To evaluate this matrix, we set $\tilde{\phi}(t)$ to be an arbitrary differentiable function, while taking the discrete-valued $V_1^+(t)$ to be of the form
\begin{equation}\label{v}
V^+_1(t)=V^+_1(0)+\int^{t}_{0}dt\,\sum_{m=1}^{\infty}\delta(t_{m}-t)\,\Delta t_m ,
\end{equation}
as a piecewise constant function with the jumps of the value at $t_{m}$ with the corresponding gaps of $\Delta_m$. Then, the physical Hamiltonian can be evaluated from \eqref{propagator}, \eqref{v} and \eqref{phys H} as
\begin{eqnarray}
\label{physical H 2}
&&\hat{\bm{H}}(t)\nonumber
\\
&=&- \underline{P}_{\tilde{\phi}}(\hat{\Omega}_2(t),{}\hat{\Omega}_3(t))\,\partial_{t}\tilde{\phi}-\hbar\sum_{m=1}^{\infty}\delta(t-t_m) \,\\
&&\times[\,{\theta}_1({}\hat{\Omega}_2(t),{}\hat{\Omega}_3(t),\,t_{m+1})-{\theta}_1({}\hat{\Omega}_2(t),{}\hat{\Omega}_3(t),\,t_m)\,]\nonumber
\\
&\equiv&\hat{\bm{H}}_{\tilde{\phi}}(t)+\hat{\bm{H}}_{V^+_1}(t),
\end{eqnarray}
which depends on respectively the specified $\tilde{\phi}(t)$ and $V^+_1(t)$. Also, note the absence of the $\widehat{P\sgn}_I(t)$ which indicates that $$\partial_t\,\widehat{\sgn}_I(t)=0$$ and the four distinct values of $\widehat{\sgn}_I(t)$ simply label the four  decoupled superselected sectors in ${\mathbb{D}_{\text{ST}}}$.

We can now look into the semiclassical limit of the dynamics through the physical Hamiltonians.  Consider a physical state $\Psi$ with the wave function $\Psi(\Omega_2(t) ,\Omega_3(t))$ having the support of a compact region $Sup_\Psi \subset \mathbb{R}^2=\{(\Omega_2 ,\Omega_3)\}$, so that
\begin{eqnarray}
\big|\Psi\big)\equiv \int_{Sup_\Psi} d^2\Omega_I(t)\,\, \Psi(\Omega_I(t) ,\widehat{\sgn}_I(t))\,\, \big|\Omega_I(t),\widehat{\sgn}_I(t)\big), \nonumber\\
\end{eqnarray}
Accordingly, we may then identify the period of time in which the $\Psi$ may behave semiclassically. This period is given by the condition $$V_1^+(t)\in I_\Psi\equiv \bigcap_{ (\Omega_2 ,\Omega_3)\in Sup_\Psi} I_{\underline{\Omega}_1(\Omega_2 ,\Omega_3) }$$. In this period, the stationary phase approximation of the physical Hamiltonian acting on $\Psi$ is valid for us to carry on the following analysis.

Due to the discrete nature of the $V^+_1(t)$, the operator $\hat{\bm{H}}_{V ^+_1}$ takes the distributional form as expected, and we want to know if proper smearing of the distribution can lead to a meaningful semiclassical limit. Thus, according to \eqref{stationary phase} and \eqref{thetapm}, given the set of values $(\Omega_2,\Omega_3, V^+_1(t))$ with $V^+_1(t)\in I_{\underline{\Omega}_1}$, we have the asymptotic expression
 \begin{align}
\label{theta}
 &\theta_1(\Omega_2,\Omega_3,t) \nonumber\\
= &\theta^+(\Omega_1,V_1)+O(\ell^{5/2}_{Pl})\,\,|_{\Omega_1=\underline{\Omega}_1(\Omega_2,\Omega_3), V_1=V^+_1(t)}\nonumber\\
 = &\frac{-\gamma}{(\xi\gamma)^{3/2}\ell^3_{\text{Pl}}}\bigg(\sqrt{\Box}\Omega_i\ln\bigl|\tan(\frac{b^+_1}{\hbar})\bigl|-\frac{b^+_1 V^+_1(t)}{\hbar}\bigg)|_{b^+_1=b^+_1(\underline{\Omega}_1, V^+_1(t))}\nonumber\\
&+\frac{\pi}{4}+O(\ell^{5/2}_{Pl}),
 \end{align}
Using the abbreviation ${b}^+_1(\Omega_2,\Omega_3,t)\equiv {b}^+_1(\underline{\Omega}_1(\Omega_2,\Omega_3),t)$, we may then insert the expression above to \eqref{physical H 2} and obtain 
\begin{align}
\label{effective H}
& \int_{t_m} ^{t_{m+1}} dt\,\hat{\bm{H}}_{V^+_1}(t)  \,\,|\Psi)\,\,\big|_{V_1^+(t_m)\in I_\Psi}\nonumber\\
=&\big[- {\Delta t_m} \,\,\partial_t V^+_1(t_m)\,\,\frac{ b^+_1(\hat{\Omega}_2(t_m),\hat{\Omega}_3(t_m),t_m)}{(\xi\gamma)^{3/2}\ell^3_{\text{Pl}}}+O(\ell^{9/2}_{Pl})\,\big]\,|\Psi),
\end{align}
where 
\begin{equation*}
\partial_t V^+_1(t_m) \equiv \frac{{V^+_1}(t_{m+1})-{V^+_1}(t_{m})}{\Delta t_m}.
\end{equation*}

By taking the lowest order terms in $\hbar$, we obtain the semiclassical effective Hamiltonian for $\Psi$, valid in the period of time with $V_1^+(t)\in I_\Psi$, as given by
\begin{eqnarray}
\label{phyh}
\bm{H}^{\text{eff}}(t)= &-\underline{P}_{\tilde{\phi}}({}\Omega_2(t),{}\Omega_3(t))\cdot\partial_t\tilde{\phi}(t) 
\nonumber\\
&{-\frac{ 4 b^+_1({}\Omega_2(t),{}\Omega_3(t),t)}{3\gamma\hbar\sqrt{\Box}}}\cdot\partial_{t}{V^+_1}(t)\nonumber\\
=&-\underline{P}_{\tilde{\phi}}({}\Omega_2(t),{}\Omega_3(t))\cdot\partial_t\tilde{\phi}(t) 
\nonumber\\
&-\frac{2\,\underbar{c}_1({}\Omega_2(t),{}\Omega_3(t),t)}{3\gamma \sqrt{p_1(t)}}\cdot\partial_{t}{V^+_1}(t)\nonumber\\
\end{eqnarray}
where the $\underbar{c}_1({}\Omega_2(t),{}\Omega_3(t),t)$ and $p_1(t)$ denote respectively the constrained value of $c_1$ and the assigned value of $p_1$. Again, beyond this period of time, the Schr\"odinger dynamics of $\Psi$ is purely quantum and has no classical limits.

\subsubsection{ Recovery of GR LQC with $\tilde{\phi}(t)={0}$ }

 Now, let us observe that the effective semi-classical Hamiltonian $\bm{H}^{\text{eff}}(t)$ recovers that of the  GR LQC with arbitrary $V^+_1(t)$ when we set $\tilde{\phi}(t)=0$.

The classical theory of GR is given by setting $\phi=e^{\tilde{\phi}}=1$ as a fixed parameter in the action \eqref{Brans-Dicke Action}, which then becomes the familiar Einstein-Hilbert action in vacuum. Without the conformal symmetry, the canonical Bianchi-I GR is governed by only the scalar constraint $C_H$ defined in \eqref{original constraints} expressed in terms of the same canonical coordinates $(c_i,p_i)$ as the reduced Ashtekar variables, with the multiplier related to the lapse function via \eqref{lapse function} -- all under the setting of $ \phi=e^{\tilde{\phi}}=1$. In the common scheme we follow, the GR LQC would be based on the loop-corrected scalar constraint $C_H^L$ given in \eqref{regularized constraints}. 

Now, the general evolutions of our conformal model in the semi-classical limits should appear as the trajectories on the constraint surface, generated by the specific combinations of the constraints in the form
\begin{equation}
N(t) C^L_{H}+\lambda(t) C^L_s
\end{equation}
where the Lagrangian multipliers $N(t)$ and $\lambda(t)$ are the time-dependent phase space functions determined by the two specified reference field values $V^+_1(t)$ and $\tilde{\phi}(t)$ via solving 
 \begin{equation}\begin{split}
\label{cond}
\{V^+_1(t)\,,\, N(t) C^L_{H}+\lambda(t) C^L_s\}=\partial_t V^+_1(t) \,\,\text{and}\\
\{\tilde{\phi}(t)\,,\, N(t) C^L_{H}+\lambda(t) C^L_s\}=\partial_t\tilde{\phi}(t)=0.
\end{split}\end{equation}
In the $\tilde{\phi}(t)={0}$ case we are concerned with, the solutions are given by
\begin{equation}\begin{split}
\label{lapse1}
N(t)&=\frac{\gamma^2}{(\Omega_2+\Omega_3)}\frac{\dot{V}^+_1}{\sqrt{V^{+\,2}_1-\Omega^2_1}} \\
\lambda(t)&=0.
\end{split}\end{equation}
With these solutions, the condition \eqref{cond} implies that
\begin{equation}
N(t) C^L_{H} \doteq \frac{2c_1}{3\gamma \sqrt{p_1}}\cdot\partial_{t}{V^+_1}(t)+h(\Omega_I,V_I, t)
\end{equation}
for a certain function $h(\Omega_I,V_I, t)$. Here the weak equality symbol $\doteq$ indicates the matching up to the first-order derivatives in the phase space variables, when evaluated on the constraint surface. In particular, the function $h$ transforming the dynamical variables must satisfy
\begin{equation}
h(\Omega_I,V_I, t)\doteq -\frac{2\,\underbar{c}_1({}\Omega_2,{}\Omega_3,t)}{3\gamma \sqrt{p_1(t)}}\cdot\partial_{t}{V^+_1}(t).
\end{equation}
It is clear now our semi-classical physical Hamiltonian $\bm{H}^{\text{eff}}(t)$ with $\partial_t\tilde{\phi}(t)=0$ is given by
\begin{equation}
\bm{H}^{\text{eff}}(t)\doteq h(\Omega_I,V_I, t) |_{\Omega_I=\Omega_I(t),\,V_I=V_I(t)},
\end{equation}
and it indeed generates the GR LQC evolution of the dynamical variables, in the assigned reference frame. 

Lastly, let us write down the semiclassical Hamilton equations of motion for the metric observables as 
\begin{equation}\begin{split}
 \frac{d}{dt}V_I(t)=&\{V_I(t),\bm{H}^{\text{eff}}(t)\}\\
 =&\sgn(\cos\frac{b_I}{2\hbar}) \frac{\Omega^2_J(t)}{(\Omega_I(t)+\Omega_J(t))^2}\sqrt{\frac{V_I^2(t)-\Omega_I^2(t)}{V_1^2(t)-\underline{\Omega}_1^2(t)}},\\
 \end{split}\end{equation}
The above result coincides with the equations of motion obtained in Ref.~\cite{martin2009physical} for the Bianchi I model of GR LQC. The predicted evolution of the universe in the large volumn regime agrees with that of GR, while in the extremely small $V_I(t)$ regimes the singularities are replaced with the ``quantum bounce" resulting from the loop corrections. Thus we conclude that, in the quantum reference frame with assigned $\tilde{\phi}(t)=0$ and $V^+_1(t)$, our conformal BD LQC model yields the GR LQC.

\subsection{Frames with specified $(P_{\tilde{\phi}}(\tau),{V}^+_1(\tau), P_S(\tau))$ for SD LQC}

For obtaining the LQC of SD, we now introduce another decomposition for the degrees of freedom in $\mathbb{K}$. Following the previous discussion in the context of our homogeneous model, we would be using the reference sector given by $\{(\hat{X}'_\mu, \hat{P}'_\mu)\} \equiv \{(\hat{\tilde\phi}, \hat{P}_{\tilde\phi}),(\hat{X}_{0\,\text{SD}}, \hat{P}_{0\,\text{SD}}),(\hat{S}, \hat{P}_{S})\}$. For easier comparison with the effective dynamics obtained in last subsection, we will choose $(\hat{X}_{0\,\text{SD}}, \hat{P}_{0\,\text{SD}})$ to be $(\hat{P}_{{\Omega}_1},\widehat{P\sgn}_1, \hat{\Omega}_1,\widehat{\sgn}_1)$ instead of following the more common York-time setting. Correspondingly, the reduced conformal constraint operator $\hat{S}$ mentioned above would be given by
\begin{eqnarray}
\label{S}
\hat{S}\equiv S(\hat{\Omega}_2, \hat{\Omega}_3)\equiv \gamma^{-1}(\hat{\underline{\Omega}}_1 +  \hat{\Omega}_2 + \hat{\Omega}_3)\,,\, 
\end{eqnarray}
which is the quantum representation of the classical reduced constraint $S=\mathcal{C}_S|_{{P}_{\tilde\phi}=0,\mathcal{C}_H=0 }$. 

Moreover, there is a canonical transformation leading to the replacement of  $(\hat{\Omega}_2, \hat{\Omega}_3)$ by the new commuting set $(\hat{S} , \hat{\mathcal{X}})$ of self-adjoint operators with
\begin{eqnarray}
\hat{\mathcal{X}}(\hat{\Omega}_2, \hat{\Omega}_3)\equiv S(-\hat{\Omega}_2, \hat{\Omega}_3).
\end{eqnarray}
One can check that $\hat{S}$ and $\hat{\mathcal{X}}$, as well as their conjugate momentum operators $$(\hat{P}_{S}, \hat{P}_{\mathcal{X}})\equiv(i\hbar\,\partial_{S}\,,\,i\hbar\,\partial_{\mathcal{X}} ),$$ all have a spectrum of $\mathbb{R}$. The spectral decomposition of $\mathbb{K}$ in the new eigenbases can be given by
\begin{align}
\label{newbasis}
&\mathbb{K}=\mathrm{span}\{\ket{\tilde{\phi},{\Omega}_1,S, \mathcal{X}, {\sgn}_I} \}\nonumber\\
&\,\,\,\,\,= \mathrm{span}\{\ket{P_{\tilde{\phi}},{P}_{{\Omega}_1},{P}_{S}, {P}_{\mathcal{X}},{P\sgn}_I}\}, \,\nonumber\\
\end{align}
with the inner product of 
\begin{align}
&\,\langle{\tilde{\phi}',\Omega'_1,S', \mathcal{X}',{\sgn}'_I|\tilde{\phi},\Omega_1,S, \mathcal{X},{\sgn}_I}\rangle\nonumber\\
&= \delta_{{\sgn}_I,{\sgn}'_I}\delta(\tilde{\phi}'-\tilde{\phi})\delta(\Omega'_1-{\Omega}_1)\delta(S'-S)\delta(\mathcal{X}'-\mathcal{X})\,.
\end{align}

Note that the above infer the relation
\begin{eqnarray}
\label{Xeigenstate}
&\ket{S, \mathcal{X}}= \Phi(\underline{\Omega}_2(S,\mathcal{X}), \underline{\Omega}_3(S,\mathcal{X}))\, \ket {\underline{\Omega}_2(S, \mathcal{X}), \underline{\Omega}_3(S,\mathcal{X})} , \, \nonumber\\
&\text{with} \nonumber\\
& \Phi^2( {\Omega}_2,{\Omega}_3) d\mathcal{X}dS= d{\Omega}_2 d{\Omega}_3\,,
\end{eqnarray}
where $\underline{\Omega}_2(S, \mathcal{X})$ and $\underline{\Omega}_3(S,\mathcal{X})$ are simply defined by the one-to-one correspondence between $(S, \mathcal{X})$ and $( {\Omega}_2,{\Omega}_3)$.

\subsubsection{ Emergence of SD LQC with ${P}_{\tilde{\phi}}(\tau)=0$ }

Following the above discussion about the reduction leading SD, we now choose $$\{\hat{T}'_\mu\}\equiv \{\hat{P}_{\tilde{\phi}}, \hat{V}^+_1, \hat{P}_S\}$$  with the assigned instantaneous values $P_{\tilde{\phi}}(\tau)\equiv 0,\,{V}^+_1(\tau),\, P_S(\tau)$ as functions of $\tau$. This leaves the dynamic degrees of freedom as 
$$\{(\hat{X}'_I, \hat{P}'_I)\}=\{\mathcal{X}, \widehat{P\sgn}_I,P_{\mathcal{X}},\widehat{\sgn}_I \}$$ and the associated eigenspace $${\mathbb{S}^\tau = \ket{{P}_{\tilde{\phi}}=0,{V}^+_1(\tau), P_S(\tau)}\otimes\mathrm{span}\{\ket{\mathcal{X},\widehat{\sgn}_I }\}}$$ for each value of $\tau$.

Similar to the previous calculations, the relevant transition amplitudes $(\mathcal{X}, {\sgn}_I,\tau)\rightarrow(\mathcal{X}',{\sgn}'_I,\tau')$ can be directly obtained. The relevant matrix elements are then given by (with the abbreviation $\Phi_{S,\mathcal{X}}\equiv\Phi(\underline{\Omega}_2(S,\mathcal{X}): \underline{\Omega}_3(S,\mathcal{X}))$):
\begin{align}
\label{SD transit1}
& \mathbb{P}_{\tau',\tau}(\mathcal{X}',{\sgn}'_I;\mathcal{X},{\sgn}_I) \nonumber\\
=&\bra{0,{V}^+_1(\tau'), P_S(\tau'), \mathcal{X}',\widehat{\sgn}'_I }\hat{\mathbb P} \ket{0,{V}^+_1(\tau), P_S(\tau), \mathcal{X},\widehat{\sgn}_I} \nonumber \\
=&\int dS \int dS'  e^{-i (S P_S(\tau)-S' P_S(\tau'))/\hbar}  \nonumber\\ 
& \times \bra{0,{V}^+_1(\tau'), S', \mathcal{X}', {\sgn}'_I}\hat{\mathbb P}  \ket{0,{V}^+_1(\tau), S, \mathcal{X},{\sgn}_I } \nonumber\\
=&\int dS' \,\Phi_{S',\mathcal{X}'}\,\Phi_{0,\mathcal{X}}\,e^{i \,S' P_S(\tau')/\hbar}\, \nonumber\\
&\times\bra{0,V^+_1(\tau'),\underline{\Omega}_2(S',\mathcal{X}'), \underline{\Omega}_3(S'\mathcal{X}'),{\sgn}'_I} \delta(\hat{\mathcal{C}}_H) \nonumber\\
&\cdot\ket{0,V^+_1(\tau),\underline{\Omega}_2(0,\mathcal{X}), \underline{\Omega}_3(0,\mathcal{X}),{\sgn}_I} \nonumber\\
=&  \gamma^2\,\alpha\,\delta_{\sgn'_I,\sgn_I}\, \delta(\mathcal{X}'- \mathcal{X})\,\, \frac{ \langle V^+_1(\tau')|\underline{\Omega}_{1}(\mathcal{X})\rangle\langle\underline{\Omega}_{1}(\mathcal{X})|V^+_1(\tau)\rangle}{|\underline{\Omega}_2(\mathcal{X})+\underline{\Omega}_3(\mathcal{X})|}\,,
\end{align}
where we use the abbreviations
\begin{eqnarray}
& \alpha\equiv \langle{{P}_{\tilde{\phi}}=0|{P}_{\tilde{\phi}}=0}\rangle=\delta(0)\,; \nonumber\\
&\underline{\Omega}_{I=2,3}(\mathcal{X})\equiv\underline{\Omega}_{I}(0,\mathcal{X})\, ; \, \underline{\Omega}_{1}(\mathcal{X})\equiv\underline{\Omega}_1(\underline{\Omega}_2(\mathcal{X}), \underline{\Omega}_3(\mathcal{X}))\,.\nonumber
\end{eqnarray}
The reader may refer to Appendix \ref{app:A} for the rigorous treatment of the distributional infinity here denoted as the $\alpha$, and as we will see in the next subsection, this infinity correctly indicates the proper distributional nature of the physical states described in this frame. The operator $\hat{\Lambda}_\tau$ for this case is then obtained as
\begin{equation}\label{lambda2}
\hat{\Lambda}_\tau=\hat{\mathbb{P}}_{\tau,\tau}^{-1/2}=\frac{\sqrt{|\underline{\Omega}_2(\hat{\mathcal{X}})+\underline{\Omega}_3(\hat{\mathcal{}X})|}}{\sqrt{ \gamma^2\,\alpha }\,|\langle V^+_1(\tau')|\underline{\Omega}_{1}(\hat{\mathcal{X}})\rangle|}\,,
\end{equation}
using which we can verify that 
\begin{eqnarray}
\label{SD transit2}
U_{\tau,\tau}=\mathbb{I} \,\,\, \text{and}\,\, \,U^{\dag}_{\tau',\tau}U_{\tau',\tau}=\mathbb{I}\,;\nonumber\\
\hat{U}_{\tau',\tau}\equiv \hat{\Lambda}_{\tau'} \hat{\mathbb{P}}_{\tau',\tau}\hat{\Lambda}_{\tau}\,.
\end{eqnarray}
 We have thus derived a Schr\"odinger theory in the domain $\mathbb{D}_{\tau}=\mathbb{D}_{\tau'}\equiv \mathbb{D}_{\text{SD}}$, described by the quantum relational observables $$\{(\hat{X}'_I(\tau), \hat{P}'_I(\tau))\}=\{\mathcal{X}(\tau), \widehat{P\sgn}_I(\tau),P_{\mathcal{X}}(\tau),\widehat{\sgn}_I(\tau) \}$$.  Again, the factor $\delta_{\sgn'_I,\sgn_I}$ in the above transition amplitudes implies  that we have $$\partial_\tau\,\widehat{\sgn}_I(\tau)=0$$ and the four distinct values of $\widehat{\sgn}_I(\tau)$ label the four decoupled superselected sectors in ${\mathbb{D}_{\text{SD}}}$ for the theory .

We may carry on to obtain and study the explicit form of the physical Hamiltonian $\hat{\bm{H}}(\tau)$ in the same way as in the GR LQC's case, although we would not do so this work. Instead, let us take a look at the special meaning of the quantum relational observables in this theory, by viewing the classical observables they represent. Recall that in the procedure to obtain the SD from the conformal BD theory, after the partial-reduction with respect to the scalar constraint $C_H$ under the conditions ${P}_{\tilde{\phi}}(\tau_1)=0$ and $ {V}^+_1= {V}^+_1(\tau_1)$ with a specific $\tau_1$, the conformal constraint $C_S$ reduces to $S$ as a phase space function in the partial-reduced phase space $(v_2, b_2,v_3, , b_3)$ at $\tau_1$. Also, by the canonical coordinate decomposition we have  
\begin{eqnarray}
\label{conf inv}
\{S ,({\mathcal{X}}, P_{\mathcal{X}})\}=0\,.
\end{eqnarray}
By the nature of any proper symplectic reduction we have mentioned, this equation in the  (partially-) reduced phase space can be ``pulled back" to the  $C_H$-constraint surface of the full phase space, leading to 
\begin{eqnarray}
\label{conf inv2}
\{C_S , ({\mathcal{X}}(\tau_1), P_{\mathcal{X}}(\tau_1))\}|_{ C_H=0}=0. 
\end{eqnarray}
It is clear now, that the quantum relational observables $\{\,\hat{\mathcal{X}}(\tau),\widehat{P\sgn}_I(\tau), \hat{P}_{\mathcal{X}}(\tau),\widehat{\sgn}_I(\tau)\,\}$ for the SD LQC should represent the classical relational observables of shapes, which do not distinguish between the BD spacetimes related by the local conformal transformations generated by $C_S$. Later, we will write down the explicit expressions of our quantum relational observables to confirm the above expectation and interpretation. Thus our quantum relational observables can be interpreted as the physical degrees of freedom in the  ``quantum shapes". Their evolution is described by our Schr\"odinger theory of shape dynamics.

\subsubsection{Equivalence to GR-LQC with maximal slicing via the linking-theory LQC}

Here, let us address two important types of transformations among the augmented quantum reference frames.

The first type of transformation corresponds to the usual reference frame transformations in GR LQC. For instance, using the quantum reference frame with the reference fields specified as $\{\hat{T}''_\mu\}\equiv\{\hat{\tilde{\phi}}, \hat{b}^+_1\equiv  \Theta(\widehat{\sgn}_1)\hat{b}_1\Theta(\widehat{\sgn}_1)\}$ with the assigned functions of $\tilde{\phi}(\sigma)=0$ and $b_1^+(\sigma)$. One can check that an Schr\"odinger theory is again obtained in this frame sharing exactly the same domain of $\mathbb{D}_{\text{GR}}$. This theory is the GR LQC model in the quantum reference frame with specified $b(\sigma)$, such that $\hat{\bm{H}}(\sigma)=\hat{H}^{GR}_{lqc}(\sigma)$. Therefore, with $\{\hat{\Omega}_I(\sigma), \widehat{\sgn}_I(\sigma) \}$ being the complete set of quantum relational observables constructed in the new reference frame, the unitary matrix $(\,\Omega_I(\sigma), {\sgn}_I(\sigma) \,|\,\Omega_I(t), {\sgn}_I(t)\, )$ here represents a usual transformation of the spacetime reference frames in GR LQC.

The second type of transformation involves the role of our model as a linking theory. Let us thus study the relation between the two frames respectively with $(P_{\tilde{\phi}}(\tau)\!=0,\,{V}^+_1(\tau), P_S(\tau))$ and $(\tilde{\phi}(t)=0,\,{V}^+_1(t))$ by computing the matrix 
\begin{align}\label{eq:matrix}
& (\,\Omega_2(t), \Omega_3(t),{\sgn}_I(t)\,|\, \mathcal{X}(\tau),{\sgn}_I(\tau)\, ) \nonumber\\
=&\frac{\delta_{{\sgn}_I(t),{\sgn}_I(\tau) }}{\alpha^{1/2}}\,e^{\frac{-i}{\hbar}\,[\,\,\theta_1(\Omega_2(t),\Omega_3(t),\,t)\,-\,\,\theta_1(\underline{\Omega_2}(\mathcal{X}(\tau)),\underline{\Omega_3}(\mathcal{X}(\tau)),\,\tau)\,\,]}\nonumber\\
&\times \, (\,\Omega_2(t), \Omega_3(t),{\sgn}_I(t)\,|\,S(t)=0,\mathcal{X}(t),{\sgn}_I(t)\,) \,,
\end{align} 
where $|\,S(t),\mathcal{X}(t),{\sgn}_I(t)\,)$ denotes the eingenstate of the GR LQC observables $\{\hat{S}(t), \hat{\mathcal{X}}(t),{\sgn}_I(t) \}$. Eq.~\eqref{eq:matrix} then implies that
\begin{eqnarray}
\label{sdnorm}
|\, \mathcal{X}(\tau),+1\, )= \alpha^{-1/2}\,|\,S(t)=0,\mathcal{X}(t), +1\,)\,|_{\tau=t,\, \mathcal{X}(\tau)=\mathcal{X}(t) }\nonumber\\
\end{eqnarray}
which means that $\mathbb{D}_{\text{SD}}\subset\mathbb{D}^*_{\text{GR}}$ and hence our SD LQC states are always given by specific distributional states in our GR LQC. More precisely, it becomes clear that the subspace spanned the shape-dynamics states in $\mathbb{D}^*_{\text{GR}}$ is precisely the eigenspace of $\hat{S}(t)$ with the eigenvalue of $S(t)=0$. Therefore, the factor of  $\alpha^{-1/2}=\delta^{-1/2}(0)$ inherited from the serving to absorb the overall infinite factor of $(S(t)=0|S(t)=0)=\delta(0)$, making the constituent SD wavefunctions $L^2$-normalizable over the remaining variable $\mathcal{X}$.

Thus, our model indeed provides the crucial feature as a quantum linking theory, in which the above transformation of the augmented quantum reference frames transforms the SD LQC into the GR LQC in the foliation satisfying $S(t)=0$. In particular, recalling that $S= V K$ with $K$ being the mean-extrinsic-curvature of the spatial hypersurface, we conclude that the SD LQC is equivalent to GR-LQC with the $K=0$ maximal slicing, as it has been commonly shown in the standard prescription of the classical Linking theory { \cite{ mercati2014shape,gielen2018gravity}}.

\section{Elementary local quantum observables and the Fadeev-Popov path integral   }

We have explored the two important Schr\"dinger theories by applying the respective quantum reference frames to our LQC model of linking theory.  Now we look into the explicit forms of their elementary relational observables and thereby show that their quantum evolutions may also be computed with the exactly defined Fadeev-Popov path integrals with quantum geometry deformations.

According to the analysis in the last section, we have the relational observables $({}\hat{\Omega}_I(t),\widehat{\sgn}_I(t), {}\hat{V}_I(t),\widehat{P\sgn}_I(t)): \mathbb{D}_{\text{GR}} \to \mathbb{D}_{\text{GR}} $ for the GR LQC. We may further use Eqs.~\eqref{rel obsv 0} and \eqref{Pi op} to express the observables explicitly in terms of the kinematical complete sets. In this case we have $\hat{\Pi}_t :\mathbb{D}_{\text{GR}}\rightarrow\mathbb{S}^t$ given by
\begin{eqnarray}
\label{quant cauchy surf}
\hat{\Pi}_t=\hat{\Lambda}_t|\tilde{\phi}(t),V^+_1(t)\bigl>\bigl<\tilde{\phi}(t),V^+_1(t)|\hat{\Lambda}_t.
\end{eqnarray}
The explicit forms of the elementary quantum relational observables can now be written down via \eqref{rel obsv 0} as
\begin{equation}\begin{split}
\label{OV}
({}\hat{\Omega}_I,{}\hat{V}_I)(t)=\hat{\mathbb{P}} \bigg(\frac{ \sqrt{|\hat{\Omega}_2+\hat{\Omega}_3|} }{\gamma|\widehat{\bigl<V^+_1|\underline{\Omega}_1\bigl>}|} \, (\hat{\Omega}_I,\hat{V}_I)\,  \frac{\gamma|\bigl<\widehat{V^+_1|\underline{\Omega}_1}\bigl>|}{ \sqrt{|\hat\Omega_2+\hat\Omega_3|}  }\bigg)\hat{\Pi}_t\,;
\end{split}\end{equation}
and the similar expression for $({}\widehat{\sgn}_I(t),{}\widehat{P\sgn}_I(t))$ is simply given by replacing $(\hat{\Omega}_I,\hat{V}_I)$ above with  $({}\widehat{\sgn}_I,{}\widehat{P\sgn}_I)$.  Now, we could look at the Dirac observable in a different way. Observe that we have the following quantum representations for the delta functions of the reference fields,
\begin{align}
\label{condition}
&\widehat{\delta({V}_1-{V}_1(t))}\equiv \frac{1}{{\Delta{V}}_1}\ket{V_1(t)}\bra{V_1(t)} ,  \nonumber\\
&\widehat{\delta(T_{\mu}-T_{\mu}(t))} \equiv \delta(\hat{\tilde{\phi}}-\tilde{\phi}(t))\cdot\Theta(\widehat{\sgn}_1)\Theta({\cos\frac{2\hat{b}_1}{\hbar}}) \nonumber\\
&\,\,\,\,\,\,\,\,\cdot\,\widehat{\delta({V}_1-{V}_1(t))}\Theta({\cos\frac{2\hat{b}_1}{\hbar}}) \Theta(\widehat{\sgn}_1) \nonumber\\
&= \frac{1}{{\Delta{V}_1}}\ket{\tilde{\phi}(t),V^+_1(t)}\bra{\tilde{\phi}(t),V^+_1(t)},
\end{align}
where $\Delta V_1= V_1(t_{m+1})-V_1(t_m)$. Furthermore, introducing the matrix $\dot{T}$ with its elements given by the Poisson brackets $\dot{T}_{\nu\mu}\equiv\{\mathcal{C}_{\nu}, T_\mu\}_{\nu=S,H}$, one may easily check using the WKB approximations that we have a quantum representation of its determinant as
\begin{eqnarray}
\label{ghost}
\widehat{\sqrt{\det{\dot{T}}}}\equiv \frac{\sqrt{\Delta{V_1}}\, \sqrt{|\hat\Omega_2+\hat\Omega_3|} }{\gamma|\bigl<{V}^+_1(t)|\underline{\hat\Omega_1}\bigl>|}=\sqrt{\Delta{V_1}}\,\hat{\Lambda}_t.
\end{eqnarray}
Combining all the above, we obtain
\begin{align}
\label{obsv1}
 &({}\hat{\Omega}_I,{}\hat{V}_I)(t) 
=\int_{-\infty}^{+\infty}d\lambda d N\,\, e^{i\lambda\hat{C}_S}
e^{i N \hat{C}_H} \bigg( \frac{ \sqrt{|\hat{\Omega}_2+\hat{\Omega}_3|} }{\gamma|\widehat{\bigl<V^+_1|\underline{\Omega}_1\bigl>}|} \nonumber\\
& (\hat{\Omega}_I,\hat{V}_I) \frac{\gamma|\bigl<\widehat{V^+_1|\underline{\Omega}_1}\bigl>|}{ \sqrt{|\hat\Omega_2+\hat\Omega_3|}  }
\widehat{\sqrt{\det{\dot{T}}}}\,\,\widehat{\delta(T_{\mu}-T_{\mu}(t))}\,\,\widehat{\sqrt{\det{\dot{T}}}}\bigg) \nonumber\\
&e^{-i\lambda\hat{C}_S}e^{-i N\hat{C}_H}.
\end{align} 
From this integral expression one can clearly see that the observables $\hat{\Omega}_I(t)$ and $\hat{V}_I(t)$ are special quantum representations of the gauge invariant phase space functions taking the values of $\hat{\Omega}_i$ and $\hat{V}_i$ at a point of a gauge orbit where $(\tilde{\phi}, {V}^+_1)=(\tilde{\phi}(t), {V}^+_1(t))$ is satisfied.

In the same manner, we now study the relational observables $(\hat{\mathcal{X}}(\tau), \hat{P}_{\mathcal{X}}(\tau))$ for our Schr\"odinger theory of SD-LQC. Similar to \eqref{OV}, they are given by the universal structure following from the algorithm as
\begin{align}
\label{XP}
(\hat{\mathcal{X}}, \hat{P}_{\mathcal{X}})(\tau) =& \hat{\mathbb{P}} \bigg(\frac{\sqrt{|\underline{\Omega}_2(\hat{\mathcal{X}})+\underline{\Omega}_3(\hat{\mathcal{X}})|}}{\sqrt{ \gamma^2\,\alpha }\,|\langle V^+_1(\tau')|\underline{\Omega}_{1}(\hat{\mathcal{X}})\rangle|} \, (\hat{\mathcal{X}}, \hat{P}_{\mathcal{X}})\, \nonumber\\
& \frac{\sqrt{ \gamma^2\,\alpha }\,|\langle V^+_1(\tau')|\underline{\Omega}_{1}(\hat{\mathcal{X}})\rangle|}{\sqrt{|\underline{\Omega}_2(\hat{\mathcal{X}})+\underline{\Omega}_3(\hat{\mathcal{X}})|}}\bigg)\hat{\Pi}_\tau\,,
\end{align}
where we have
\begin{align}
\label{quant cauchy surf2}
\hat{\Pi}_\tau=\hat{\Lambda}_\tau|{P}_{\tilde{\phi}}(\tau),V^+_1(\tau), P_S(\tau)\bigl>\bigl<{P}_{\tilde{\phi}}(\tau),V^+_1(\tau), P_S(\tau)|\hat{\Lambda}_\tau. 
\end{align}
Also, analogous to the previous case we may set
\begin{align}  
\label{condition2}
& \widehat{\delta(T'_{\mu}-T'_{\mu}(t))}  \nonumber\\
\equiv& \delta(\hat{P}_S-P_S(\tau) ) \delta(\hat{P}_{\tilde{\phi}}-{P}_{\tilde{\phi}}(\tau))\nonumber\\
&\cdot\Theta(\widehat{\sgn}_1)\Theta({\cos\frac{2\hat{b}_1}{\hbar}})  \widehat{\delta({V}_1-{V}_1(t))} \Theta({\cos\frac{2\hat{b}_1}{\hbar}})\Theta(\widehat{\sgn}_1)\nonumber\\
=& \frac{1}{{\Delta{V}_1}}|{P}_{\tilde{\phi}}(\tau),V^+_1(\tau), P_S(\tau)\bigl>\bigl<{P}_{\tilde{\phi}}(\tau),V^+_1(\tau), P_S(\tau)|.
\end{align}  
In this case, we then look for the quantum representation of the determinant of the matrix $\dot{T}'_{\nu\mu}\equiv\{\mathcal{C}_{\nu}, T'_\mu\}_{\nu=S,H\,;\,\, \mu=2,3}$ with $\{T'_2, T'_3\}\equiv\{V^+_1, P_S \}$, and find that
\begin{align}
\label{ghost2}
\widehat{\sqrt{\det{\dot{T}'}}}\equiv \frac{\sqrt{\Delta{V_1}}\, \sqrt{|\underline{\Omega}_2(\hat{\mathcal{X}})+\underline{\Omega}_3(\hat{\mathcal{X}})|} }{\gamma|\bigl<{V}^+_1(\tau)|\underline{\Omega}_1(\hat{\mathcal{X}})\bigl>|}=\sqrt{\Delta{V_1}\,\alpha}\,\hat{\Lambda}_\tau\,.
\end{align}
Combining the above results, we obtain
\begin{align}
\label{obsv2}
&(\hat{\mathcal{X}}, \hat{P}_{\mathcal{X}})(\tau) =\alpha^{-1}\int_{-\infty}^{+\infty}d\lambda d N\,\, e^{i\lambda\hat{C}_S}
e^{i N \hat{C}_H} \nonumber\\
& \bigg( \frac{ \sqrt{|\underline{\Omega}_2(\hat{\mathcal{X}})+\underline{\Omega}_3(\hat{\mathcal{X}})|} }{\gamma|\bigl<{V}^+_1(\tau)|\underline{\Omega}_1(\hat{\mathcal{X}})\bigl>|}   (\hat{\mathcal{X}}, \hat{P}_{\mathcal{X}}) \frac{ \sqrt{|\underline{\Omega}_2(\hat{\mathcal{X}})+\underline{\Omega}_3(\hat{\mathcal{X}})|} }{\gamma|\bigl<{V}^+_1(\tau)|\underline{\Omega}_1(\hat{\mathcal{X}})\bigl>|} \nonumber\\
&\widehat{\sqrt{\det{\dot{T}'}}}\,\,\widehat{\delta(T'_{\mu}-T'_{\mu}(t))} \widehat{\sqrt{\det{\dot{T}'}}}\bigg)e^{-i\lambda\hat{C}_S}e^{-i N\hat{C}_H}.
\end{align}
From this integral expression, one can clearly see that the observables $\hat{\mathcal{X}}(\tau)$ and $\hat{P}_{\mathcal{X}}(\tau)$ are specific quantum representations of the gauge invariant phase space functions, taking the values of $\mathcal{X}$ and ${P}_{\mathcal{X}}$ at a point of a gauge orbit where the condition $(P_{\tilde{\phi}}, {V}^+_1, P_S)=(0, {V}^+_1(\tau), P_S(\tau))$ is satisfied. Thus indeed they carry the meaning of the instantaneous shape degrees of freedoms $\mathcal{X}$ and ${P}_{\mathcal{X}}$ in relation to $V_1$, for the selected physical states satisfying $P_{\tilde{\phi}}=0$. Note that the constant factor $\alpha^{-1}$ correctly cancels out the infinity from the factor of $\delta(P_{\tilde{\phi}}-0)$ in $\delta(T'_{\mu}-T'_{\mu}(t))$, which in this homogeneous model serves only to superselect the physical states.

With the above affirmations in the meaning of our quantum observables, we may now look for the corresponding Fadeev-Popov path-integral expression for the Schr\"odinger propagators we derived previously. Recall that the Schr\"odinger propagator in the physical time $t$ is obtained from the corresponding Wheeler-Dewitt propagator $\mathbb{P}_{t',t}$ through a transformation at the initial and final time by the operator $\hat{\Lambda}$. Already suggested by the form of the relational observables above, this path integral for the Schr\"odinger propagator should be given by $\mathbb{P}_{t',t}$ as a Fadeev-Popov path integral with a quantum deformed effective BD action, with certain additional boundary factors contributed by the operator $\hat{\Lambda}$. Thus our formulation generates an exact formula for such a path integral. Using the property $\hat{\mathbb{P}}\hat{\Pi}_{t_{i}}=\hat{\mathbb{I}}: \mathbb{D}_{GR}\to\mathbb{D}_{GR}$ of the quantum Cauchy surfaces, we have
\begin{align}\label{fp path int op}
&(\,b'_I(t_n)\,|\,b_I(t_0)\,) \nonumber\\
=& \bra{ T_{\mu}(t_n), b'_I}\hat{\Lambda}_{t_n}\,\hat{\mathbb{P}}\hat{\Pi}_{t_{n-1}}\hat{\mathbb{P}}....\,\hat{\Pi}_{t_{2}}\hat{\mathbb{P}}\hat{\Pi}_{t_{1}}\mathbb{P}\, \hat{\Lambda}_{t_0}\,\ket{T_{\mu}(t_0),b_I }\,, \nonumber\\
\end{align}
which is the operator form of the boundary-transformed Fadeev-Popov path integral. Indeed, by inserting the expressions \eqref{quant cauchy surf}, \eqref{condition},  \eqref{ghost}, \eqref{rigging map} and \eqref{b modes}, Eq.~\eqref{fp path int op} can be calculated as
\begin{align}
\label{fp path int}
&\lim_{n\to\infty} (\,b_I(t_n)\,|\,b_I(t_0)\,) \nonumber\\
=&\int^{b'_I(t_n)}_{b_I(t_0)}  DN\, \,D\lambda\,D\tilde{\phi}\,DP_{\tilde{\phi}}\,D^{L}v_i\,D^{L}b_i\,\,\nonumber\\
& \sqrt{\det{\dot{T}}(t_n)}\sqrt{\det{\dot{T}}(t_0)}\,\,\det{\dot{T}}\,\delta(T_{\mu}-\bar{T}_{\mu})\, \nonumber\\
& \times \exp{-i\int_{t_0}^{t_n}dt \,(P_{\tilde{\phi}}\partial_t\tilde{\phi}+\,b_i \partial_t v_i\,-N \mathcal{C}^{L}_H+ \lambda\mathcal{C}^{L}_S\, +\eta\,)}.
\end{align}
The expression \eqref{fp path int} shows a deformed Fadeev-Popov path integral with the additional boundary transformation factors $(\det{\dot{T}})^{1/2}(t_n)$ and $(\det{\dot{T}})^{1/2}(t_0)$. This integral from the loop quantization is given by correcting the original Fadeev-Popov path integral based on the direct Fock quantization of the original constraint system \eqref{original constraints},  by specific quantum geometric deformations happening in the following two places. First, the measure $D^{L}v\, D^{L}b$ of the paths are now precisely derived from the quantum kinematics that we have introduced, and it is different from the usual $Dv\, Db$ due to the distinct loop quantum kinematics. Particularly, the measure $D^{L}v_i$ should reflect the discrete property of the $v_i$, while correspondingly the measure $D^{L}b_i$ should give a cut-off so that they become zero outside the range $[-\pi\hbar, \pi\hbar]$ of the quantum spectra. Second, the effective constraints should be given by the loop regularized constraints $\mathcal{C}^L_H$ and $\mathcal{C}^L_S$ plus the additional corrections $\eta=O(\hbar)$ from the normal ordering of the non-commuting operators. Furthermore, through the path integral expression, we clearly see that $\hat\Lambda$ is an operator representation of the square root of the Fadeev-Popov ghost contribution, and indeed the above boundary transformations by such factors have been shown to convert a gauge invariant Fadeev-Popov path integral into the corresponding reduced phase space path integral based on the usual Hamiltonian theory in the given physical time.

Similarly for the Sch\"odinger propagator of SD, by the property $\hat{\mathbb{P}}\hat{\Pi}_{\tau_{i}}=\hat{\mathbb{I}}: \mathbb{D}_{SD}\to\mathbb{D}_{SD}$ of the quantum Cauchy surfaces we have
\begin{align}
\label{fp path int op2}
&(\,{\mathcal{X}'}(\tau_n),+1\,|\,{\mathcal{X}}(\tau_0),+1\,) \nonumber\\
=&\bra{T'_{\mu}(\tau_n),{\mathcal{X}'}, +1} \hat{\Lambda}_{\tau_n}\,\hat{\mathbb{P}}\hat{\Pi}_{\tau_{n-1}}\hat{\mathbb{P}}....\,\hat{\Pi}_{\tau_{2}}\hat{\mathbb{P}}\hat{\Pi}_{\tau_{1}}\mathbb{P}\, \hat{\Lambda}_{\tau_0} \nonumber\\
&\ket{T'_{\mu}(\tau_0),{\mathcal{X}}, +1}\,,
\end{align}
which again can be written into the path integral form of 
\begin{align}
\label{fp path int2}
&\lim_{n\to\infty}(\,{\mathcal{X}'}(\tau_n),+1\,|\,{\mathcal{X}}(\tau_0),+1\,) \nonumber\\
=&\alpha^{-1}\int^{\mathcal{X}'(\tau_n),+1}_{\mathcal{X}(\tau_0),+1}  DN\, \,D\lambda\,D\tilde{\phi}\,DP_{\tilde{\phi}}\,D^{L}v_i\,D^{L}b_i\,\,\,\nonumber
\\
&  \sqrt{\det{\dot{T}'}(\tau_n)}\sqrt{\det{\dot{T}'}(\tau_0)}\,\,  \det{\dot{T}'}\,\delta(T'_{\mu}-\bar{T}'_{\mu}) \times \, \nonumber\\
&\exp{-i\int_{\tau_0}^{\tau_n}d\tau \,(P_{\tilde{\phi}}\partial_{\tau}\tilde{\phi}+\,b_i \partial_{\tau} v_i\,-N \mathcal{C}^{L}_H+ \lambda\mathcal{C}^{L}_S\, +\eta\,)}.
\end{align}
Here the factor of $\alpha^{-1}$ is consistent with the normalization condition given in Eq.\eqref{sdnorm}. Under the respective quantum reference frames, the quantum gauge symmetries corresponding to the constraints $\{\hat{C}_S, \hat{C}_H\}$ are hidden (or ``broken") for the two Schr\"odinger theories. 

The expressions \eqref{fp path int} and \eqref{fp path int2} allow us to translate our quantum reference frame approach under the canonical formulation into the path integral framework. Note that those integrals are generated by the fundamental Schr\"odinger propagators of the LQC model, and they are well defined with the naturally given regulators coming from the underlying quantum geometry of the canonical theory.

\section{Summary and Discussion} 

In this work, we have constructed the LQC model of conformal BD theory in the Bianchi-I setting and explored its contents of physical dynamics and the quantum geometric linking theory. 

This model is governed by the quantum scalar constraints and quantum conformal constraints, both deformed by the input of the quantum geometry characteristic of LQG, with the solutions of physical states representing the quantum spacetimes with the additional (loop corrected) conformal symmetries. Importantly, we have used the rigging map operator $\hat{\mathbb P}$ to construct the physical Hilbert space $\mathbb{H}$ for the model, and this allows us to use the quantum reference frame approach to construct the elementary quantum relational observables—describing the chosen dynamical sector with respect to the specified remaining reference sector, while manifestly maintaining the full interactions and symmetries of the system. Remarkably, applied to the conformal BD model, the quantum reference frames naturally acquire the interpretation of the desired augmented frames, allowing us to specify both the moment of time and the conformal scales for the associated via a proper reference sector. By construction, all the instantaneous quantum relational observables associated with a given frame share the same domain of physical states $\mathbb{D}\tilde\subset\mathbb{H}$; preserving the elementary algebra \eqref{obsv alge} in $\mathbb{K}$, they provide the elementary Heisenberg observables for the Schr\"odinger cosmic evolutions of the physical states of quantum spacetimes in $\mathbb{D}$. 

Using the frame with the specified $(\tilde{\phi}(t),{V}^+_1(t))$, we derived the Schr\"odinger theory describing the physical states in $\mathbb{D}_{GR}\tilde\subset\mathbb{H}$. Specifically, the known GR LQC dynamics is recovered using the frame with $\tilde{\phi}(t)=0$. In the second frame with the specified $(P_{\tilde{\phi}}(\tau)=0,{V}^+_1(\tau), P_S(\tau))$, we derived the second Schr\"odinger theory, which turns out to be the SD LQC describing the physical states in $\mathbb{D}_{SD}\tilde\subset\mathbb{H}$. Further, via physical Hilbert space $\mathbb{H}$, we obtain the transformation matrix between the two wave-function representations from the two theories, for the physical states in the overlapping domain $\mathbb{D}_{SD}\cap \mathbb{D}^*_{GR}$. As a result, the wave functions of SD LQC transform into the special wave functions of GR LQC with zero mean curvature $K=0$. While reproducing the result of the linking theory in the classical and continuous limit, this relation between the two LQC models incorporates the new ingredients of the quantum geometry of LQG. In the next step upon the base of this work, it is possible for us to look into the deeper principles of a quantum linking theory, such as the symmetry-trading mechanism happening at the quantum level. In this direction of research, we may ultimately explore the possibility of having the distinct full LQG theories emerging from an underlying quantum linking theory, under the various augmented quantum reference frames. 

Our results also provide a concrete framework for studying the scaling behavior of LQC linking theory. If we treat our conformal BD LQC as a Planck-scale cosmic model, the subject may be studied via two approaches associated with either side of Eqs. \eqref{fp path int} and \eqref{fp path int2} based on our construction. From the canonical view, we may study the dynamics of the ``low-energy observables" obtained from our elementary local quantum observables, through the proper coarse-graining procedure involving the time-averaging to the given scales. In this way, the propagator of the associated effective theory would be given by replacing the eigenbasis in the left-hand sides of Eqs. \eqref{fp path int} and \eqref{fp path int2}, with the eigenbasis of the low-energy observables. On the right-hand side, we expect the corresponding smearing effect to manifest through the running effective action of the path integral, which may be studied in the more familiar Wilson’s renormalization approach. Hopefully, we will be able to obtain the concrete low-energy limits of our cosmological model, which would shed light on conformal symmetry’s possible roles in LQG, especially in regard to its predictions in the observed scales of the current universe.

\section*{Acknowledgement}
We acknowledge Qian Chen for useful discussions in the early stage of this work. This work is supported by the National Natural Science Foundation of China with Grant No.11875006, No. 11961131013 and No. 12275022.

\appendix

\section{Treatment behind the $\alpha$ factor in the case of SD LQC}\label{app:A}

{For the case of SD LQC, here we show that our intuitive usage of $\alpha \equiv \delta(0)$ for the distributive physical states gives the correct results identical to the ones from the following rigorous treatment.  }

{ First, we treat the eigenstate $\ket{P_{\tilde\phi}=0}$ as the limit of the regularized form given by
\begin{equation}\label{A0}
\ket{P_{\tilde \phi}=0}\equiv \lim_{\rho\to 0}\ket{P^{\rho}_{\tilde \phi}=0}\equiv \lim_{\rho\to 0}\frac{1}{\rho} \int_{-\rho/2}^{\rho/2}\varphi(P_{\tilde \phi}/\rho)\ket{P_{\tilde \phi}} \,dP_{\tilde \phi}
\end{equation}
where the $\varphi(x)$ satisfying $$ \int_{-1/2}^{1/2}\varphi(x)\,dx=1$$ is any real function suitable for the above regularization of the Dirac-delta distribution. This way, we are viewing our chosen quantum reference frame as the limit of a sequence of distinct ``regularized quantum reference frames" labeled by $\rho$ and assigned with $\mathbb{S}^\tau_{\rho}$.  Next, one may simply carry out the steps of calculation from \eqref{SD transit1} to \eqref{SD transit2} in the regularized frame before taking the limit $\rho\to 0$.  Specifically, by evaluating the matrix elements of  $\hat{\mathbb P}$ as in \eqref{SD transit1} using instead the basis of $\mathbb{S}^\tau_{\rho}$ above, one obtains the corresponding relevant transition amplitudes $\mathbb{P}^\rho_{\tau',\tau}$ given by (with the abbreviation $|{P}^{\rho}_{\tilde{\phi}}=0\rangle\equiv |0_\rho\rangle$)
 \begin{align}\label{A2}
 &\mathbb{P}^\rho_{\tau',\tau}(\mathcal{X}',{\sgn}'_I;\mathcal{X},{\sgn}_I) \nonumber\\
=&\bra{0_\rho,{V}^+_1(\tau'), P_S(\tau'), \mathcal{X}',\widehat{\sgn}'_I }\hat{\mathbb P} \ket{0_\rho,{V}^+_1(\tau), P_S(\tau), \mathcal{X},\widehat{\sgn}_I} \nonumber \\
=&  \frac{ \gamma^2}{\rho}\,\delta_{\sgn'_I,\sgn_I}\, \delta(\mathcal{X}', \mathcal{X})\,\, \frac{ \langle V^+_1(\tau')|\underline{\Omega}_{1}(\mathcal{X})\rangle\langle\underline{\Omega}_{1}(\mathcal{X})|V^+_1(\tau)\rangle}{|\underline{\Omega}_2(\mathcal{X})+\underline{\Omega}_3(\mathcal{X})|}\,\nonumber\\
&+O(\rho).
\end{align}
Due to the smeared form of the $\ket{0_\rho}$ given in \eqref{A0}, the above calculation involves the integrations in $P_\phi$ over the interval $(-\rho/2, \rho/2)$. In the last equality, we simply Taylor- expand the integrand at $P_\phi=0$ over the interval and obtain the explicit leading-order term of $O(P_\phi^0)\sim O(\rho^0)$. From this result, the associated $\hat{\Lambda}^\rho_\tau$ can be easily checked to be
\begin{equation}\label{A3}
\hat{\Lambda}^\rho_\tau=\hat{\mathbb{P}}_{\tau,\tau}^{\rho\,-1/2}=\frac{\sqrt{\rho\,|\underline{\Omega}_2(\hat{\mathcal{X}})+\underline{\Omega}_3(\hat{\mathcal{X}})|}}{\sqrt{ \gamma^2 }\,|\langle V^+_1(\tau)|\underline{\Omega}_{1}(\hat{\mathcal{X}})\rangle|}+O(\rho^2)\,,
\end{equation}
and finally we obtain the Schr\"odinger propagator in our chosen frame, by taking the limit of the propagators in the regularized frames as 
\begin{equation}\label{A4}
\hat{U}_{\tau',\tau}\equiv \lim_{\rho\to 0} \hat{U}^\rho_{\tau',\tau}= \lim_{\rho\to 0}\hat{\Lambda}^\rho_{\tau'} \hat{\mathbb{P}}^\rho_{\tau',\tau}\hat{\Lambda}^\rho_{\tau}. 
\end{equation}
By comparing \eqref{A2},  \eqref{A3} and \eqref{A4} with \eqref{SD transit1}, \eqref{lambda2} and \eqref{SD transit2}, we can now interpret the distributive factor $\alpha$ as the short-hand for 
$$\alpha^\rho\equiv 1/\rho=\delta^\rho(0),$$ before taking the limit $\rho\to 0$ in the end. Further, it is clear that when taking the limit upon the above propagator, all the pre-factors with $\rho$ cancel out each other completely, and this provides the precise meaning of the ``canceling out of the $\alpha$ factors" stated earlier.  Due to such cancelation, the limit in \eqref{A4} indeed converges to the same propagator we obtained in \eqref{SD transit2} using the shorthand expressions \eqref{lambda2} and \eqref{SD transit2}. Finally, all of these key results are independent of the explicit form of the regularization function $\varphi(x)$, just as how they should be.}

%


\end{document}